\documentclass[10pt,conference]{IEEEtran}

\AtBeginDocument{%
  \providecommand\BibTeX{{%
    \normalfont B\kern-0.5em{\scshape i\kern-0.25em b}\kern-0.8em\TeX}}}

\usepackage{booktabs}
\usepackage{ifthen}
\usepackage{color}
\usepackage{url}
\usepackage{amsmath}
\usepackage{amssymb}
\usepackage{xspace}
\usepackage{framed}
\usepackage[T1]{fontenc}
\usepackage{enumerate}
\usepackage[linesnumbered,ruled,vlined]{algorithm2e}
\usepackage{array,multirow,graphicx}
\usepackage{float}
\usepackage{balance}
\usepackage{tikz}
\usepackage{calc}
\usepackage{subfigure}
\usepackage{listings,amsfonts}
\usepackage{amsmath,xcolor,pifont}
\usepackage{url}
\usepackage{xspace}
\usepackage{endnotes}
\usepackage{hyperref}
\usepackage{multirow,makecell}
\usepackage[misc]{ifsym}

\definecolor{yellow}{RGB}{255,255,153}
\definecolor{grey}{RGB}{224,224,224}

\newcommand{\haoyu}[1]{{\color{red} HW: #1}}

\newcommand{\richard}[1]{{\color{blue} RL: #1}}

\newcommand{\para}[1]{\smallskip\noindent{\bf {#1}. }}
\newcommand{\tabincell}[2]{\begin{tabular}{@{}#1@{}}#2\end{tabular}}

\begin{document}
\title{
When the Open Source Community Meets COVID-19:
Characterizing COVID-19 themed GitHub Repositories}

\author{
\IEEEauthorblockN{Liu Wang$^{1}$, Ruiqing Li$^{2}$, Jiaxin Zhu$^{3}$, Guangdong Bai$^{2}$, Haoyu Wang$^{1}$}
\IEEEauthorblockA{
$^{1}$ Beijing University of Posts and Telecommunications, Beijing, China\\
$^{2}$ The University of Queensland, Australia
$^{3}$ Institute of Software, Chinese Academy of Sciences, China
}
}

\maketitle
\pagestyle{plain}

\begin{abstract}
Ever since the beginning of the outbreak of the COVID-19 pandemic, researchers from interdisciplinary domains have worked together to fight against the crisis. The open source community, plays a vital role in coping with the pandemic which is inherently a collaborative process. Plenty of COVID-19 related datasets, tools, software, deep learning models, are created and shared in research communities with great efforts. However, COVID-19 themed open source projects have not been systematically studied, and we are still unaware how the open source community helps combat COVID-19 in practice. To fill this void, in this paper, we take the first step to study COVID-19 themed repositories in GitHub, one of the most popular collaborative platforms. We have collected over 67K COVID-19 themed GitHub repositories till July 2020. We then characterize them from a number of aspects and classify them into six categories. We further investigate the contribution patterns of the contributors, and development and maintenance patterns of the repositories. This study sheds light on the promising direction of adopting open source technologies and resources to rapidly tackle the worldwide public health emergency in practice, and reveals existing challenges for improvement.

\end{abstract}

\maketitle

\section{Introduction}

The COVID-19 pandemic has quickly become a world-wide crisis. Since its outbreak, COVID-19 has attracted great attention from various research communities. Researchers from interdisciplinary domains work together to fight against the crisis. Beyond the medical domain, computer scientists have adopted advanced information techniques like machine learning to help medical practitioners deal with COVID-19~\cite{shan2020lung,barstugan2020coronavirus,NOUR2020106580}. 
\textit{The open source community plays a vital role in coping with the pandemic which is inherently a collaborative process}. Existing research efforts necessitate that the datasets used for the studies should be open sourced to promote the extension and collaboration of the work in the fight against COVID-19. Thus, in the early stage of COVID-19, plenty of datasets including the statistics of COVID-19 cases, the COVID-19 diagnosis datasets from X-ray images and cough sounds, and COVID-19 emotional and sentiment datasets from social media, are created and shared in our research communities with great efforts~\cite{shuja2020covid-19, Chen_2020}. Beyond the datasets, the open source community has shared a large number of tools (e.g., online tracker), software (e.g., contact tracing mobile apps), deep learning models (e.g., diagnose models and prediction models) to help combat COVID-19. 

GitHub, as the most popular collaborative platform in the open source community, has gained great attention during this pandemic. Most of the COVID-19 datasets and software are open sourced on it.
By the time of this study, there are over 67K COVID-19 themed GitHub repositories. 
Although plenty of research studies in the software engineering community have analyzed GitHub from different perspectives~\cite{IJoC6136, 8255005, lemay:OASIcs:2019:10197}, to the best of our knowledge, the COVID-19 themed GitHub repositories have not been systematically characterized. 
We are still unaware \textit{how the open source community helps combat COVID-19 in practice}, and there remain a number of interesting directions to explore, e.g., the popularity and trends of COVID-19 themed repositories, the characteristics of contributors, the features and categories of COVID-19 repositories, and the development and maintenance behaviors.

\textbf{This Work.}
In this paper, we present the first large-scale empirical study of COVID-19 themed repositories on GitHub.
We first make efforts to harvest a comprehensive dateset of COVID-19 themed GitHub repositories. By the time of July 17, 2020, we have collected 67,079 repositories in total (\textbf{See Section~\ref{sec:dataset}}).
Leveraging the dataset, we perform a systematic analysis including popularity and trends analysis, contributor characterization, and national responsiveness analysis (\textbf{See Section~\ref{sec:general}}).
Then, we take advantage of natural language processing (NLP) techniques to classify these repositories into six major categories, including data, contact tracing, toolkit, forecast \& simulation, detection \& diagnosis, and helpful in some ways (\textbf{See Section~\ref{sec:classification}}).
We further perform an analysis to understand the development and maintenance behaviors of COVID-19 themed repositories (\textbf{See Section~\ref{sec:behavior}}).
Our work is the first comprehensive study to demystify the reaction of open source community to the pandemic. Our exploration gives a first impression on the landscape of the COVID-19 themed GitHub repositories, revealing some interesting observations:

\begin{itemize}
    \item \textbf{COVID-19 themed repositories are prevalent in GitHub.} 
    We have identified over 67K COVID-19 themed GitHub repositories contributed by around 68K unique contributors. Most repositories were created after March 11th, the time when COVID-19 outbreak was officially declared a pandemic by WHO. As the disease spread widely, the number of related repositories in GitHub immediately and significantly increased, \textit{demonstrating the open source community's rapid response to the pandemic}.
    \item \textbf{The aims of COVID-19 themed repositories cover various aspects.} We found a wide diversity of repositories and constructed a taxonomy to classify them into six categories. 
    Repositories are unevenly distributed across categories. 
    Subtle differences arise when looking in detail on a per-category basis with regard to the boom time.
    \item \textbf{Most COVID-19 repositories are developed rapidly, while the maintenance lifecycles are short-lived.
    } The development process for most repositories is rapid and intense in the early stages after creation.
    However, most repositories are not well maintained over the long run, and soon become inactive. 
    Besides, the activity of contributors are not significantly affected by the lockdown of cities.
\end{itemize}

Our results highlight the practical and potential value of open source technologies and resources in handling global crisis.
They also imply that the advanced techniques and mechanisms of popularization, internationalization, data and software sharing, contribution gathering, etc., are required for more effective collaboration of open source community in global emergencies.
To boost future research, we have released the collected dataset to the research community at \url{https://covid19-repos.github.io}.

\section{Background and Related Work}
\label{sec:background}

\subsection{GitHub Repositories}

GitHub is one of the most popular collaborative platforms primarily used for software development.
There have been millions of open source repositories in GitHub so far,
which cover various topics including different languages, frameworks, applications, events, etc.
Among them, COVID-19 studied in this paper is a very special one in 2020, and even in the history of GitHub.
The collaboration in GitHub is based on the \textit{git} version control system. 
Developers can create their own copies of a repository (aka fork) and make local changes.
The \textit{git} commits record what and when they modify.
To contribute back to the repository,
developers request the maintainers to pull the code from their forks (aka pull-request).
Maintainers can collaborate with others to review the pull-request and decide to merge or not.
The developers of the merged code are contributors of the repository.
GitHub also integrates some social features.
Users are able to \textit{follow} other users, \textit{watch} and \textit{star} repositories.
Users also have profiles of identifying information, e.g., their locations and companies.

\subsection{Analyzing GitHub Repositories}
Due to its popularity, social features, and availability of the rich data,
GitHub has attracted great attention from the software engineering community. 
Some studies have been conducted to understand GitHub repositories.
Kalliamvakou et al.~\cite{kalliamvakou2014promises} highlighted important characteristics of the GitHub repositories,
e.g., most repositories have very few commits and are inactive, 
and a large portion of repositories are not for software development.
Meanwhile, techniques were proposed to classify the GitHub repositories~\cite{zhang2019higitclass}, commits~\cite{casalnuovo2017gitcproc} and README~\cite{PranaTTAL19}. 

As collaboration is an inherent feature of software engineering projects, many studies take use of GitHub to investigate 
efficient collaboration models. 
Thanks to the rich data GitHub provides, many insights have been identified. 
Studies show that the pull-based collaboration model is more effective~\cite{zhu2016effectiveness},
although it faces a number of challenges~\cite{gousios2016pull}.
Factors which impact the integration of pull-requests are studied from many perspectives~\cite{zou2019how,KononenkoRBGTW18,ValeSSAA20}.
Pull-request reviewers are recommended~\cite{SoaresJPM18,YuWYL14}, and team structures are studied~\cite{MezouarZZ19} to improve the efficiency.
The code and development techniques in GitHub are also analyzed.
Cross-project code clones~\cite{GharehyazieRKZH19} and code quality~\cite{RayPDF17} are studied.
The performance of test case prioritization techniques~\cite{LuoMZP19} and test driven development~\cite{BorleFSGH18} are evaluated.
These studies have greatly enriched our understanding of the software development process in open source community.
In this paper, we intend to have an overview of the GitHub repositories from similar aspects in the context of COVID-19, and obtain implications of tackling the large-scale emergencies from the perspective of open source community.

\subsection{Software Engineering and COVID-19}

A growing number of studies targeting COVID-19 have sprung in the community of software engineering. 
Neto et al. ~\cite{neto2020deep} investigated the impact of COVID-19 on software projects and software development professionals through mining software repositories and a survey study. 
A few studies seek to understand developer productivity at technology companies due to the almost overnight migration to working at home for software developers. For example, Ford et al. ~\cite{ford2020tale} presented a survey study to understand benefits/challenges since working from home and analyze factors that affect developer productivity over time. Ralph et al.~\cite{Ralph_2020} used a survey to
understand the effects of the pandemic on developers' well-being and productivity, and discussed how organizations may better support their employed developers.
Bao et al.~\cite{bao2020does} conducted a case study on Baidu Inc. to understand both positive and negative impacts of working form home on developer productivity. 
To the best of our knowledge, no previous studies have systematically characterized the COVID-19 themed open source repositories on GitHub.

\section{Study Design}
\label{sec:studydesign}

We present the details of our characterization on COVID-19 themed GitHub repositories. We first describe our research questions, and then present the dataset used for our study.

\subsection{Research Questions}
Our study is driven by the following research questions:

\begin{itemize}
    \item[RQ1] \textbf{Popularity and Trends.} 
    \textit{How many COVID-19 themed repositories are there in GitHub and when are they created? How do their creations and alterations correlate with the outbreak and evolvement of the pandemic?}  
    It is interesting to investigate when the open source community reacts to the pandemic, and whether it follows similar trends with the evolvement of COVID-19.

    \item[RQ2] \textbf{Aims of COVID-19 themed Repositories.} Considering there are a large number of open source COVID-19 repositories on GitHub, it is unknown to us the aims (categories) of these repositories, i.e., \textit{what are they focused on and how are are they distributed across categories? 
    }
    
    \item[RQ3] \textbf{Development and Maintenance Behaviors.}
    Since many countries went into lockdown, we are curious \textit{whether the contributors' activities have been affected by the quarantine?} Due to the urgent nature of COVID-19, we want to study \textit{
    the development and maintenance behaviors of these repositories, i.e., are they developed quickly and well maintained?} 
\end{itemize}

\subsection{Data Collection}
\label{sec:dataset}

\subsubsection{Method}
We first need to harvest a comprehensive dataset of COVID-19 themed repositories. Our collection process is based on the official GitHub Search API\footnote{https://docs.github.com/en/rest/reference/search}, which is designed to facilitate a wide search for the public repositories and retrieve meta data about them (e.g., stars, forks, etc.). 

\para{\textbf{Collecting the related repositories}}
We first rely on a keyword matching based method to collect 
COVID-19 themed repositories. We used three search keywords (i.e., ``COVID-19'', ``COVID19'' and ``coronavirus'') as parameters to construct search queries respectively, such as ``\url{https://api.github.com/search/repositories?q=covid-19}'', and we can get a few pages of search results that best match ``covid-19'' (The query is case-insensitive). 
Since the GitHub Search API limitation of 1,000 results for each search, we have adopted a segmented approach, i.e., narrow the results of one query using search qualifiers and make multiple queries, and finally integrate them. 
We used the repository creation time (i.e., \emph{created} qualifier) and the number of stars (i.e., \emph{stars} qualifier) as the primary and secondary qualifiers respectively. 
In this way, we have collected 60,591 results for ``covid-19'', 26,289 results for ``covid19'' and 11,116 results for ``coronavirus'', respectively. 
After the de-duplication process, we obtained a dataset that contains 68,269 unique repositories. 
Besides the meta information (e.g., id, owner, creation date, stars, programming language) of each repository, we also collected profiles of its corresponding contributors, including their names, ids, locations, etc.   
We crawled these through the official API\footnote{https://api.github.com/repos/\{repo\}/contributors}.

\para{\textbf{Filtering the dataset}}
Our keyword-based collection is to match a keyword across the entire repository, such that wherever the keyword appears in a repository, we include it in our dataset.
This may cause false positives. For example, a number of repositories that are irrelevant but with description like ``[TEMPORARILY SUSPENDED because of COVID-19]'' would appear in our search results. 
We therefore conducted the following filtering process. 1) We started with checking the repository name. If the full name of the repository contains the COVID-19 wordings, e.g., covid-19, coronavirus, 2019-ncov, sars-cov-2, etc., and some anagrams of COVID-19, e.g. codiv19, condiv19, cord19, etc., we would assume this repository meets our need and keep it. Otherwise, it needs to be further examined. After this step, around 7,000 repositories are picked for inspection in the next step. 
2) For the repositories without a visible sign in their names, we turned to their descriptions. If the repository description contains no COVID-19 wordings, it is reasonable for us to consider it trivial to our research and screen it out. Otherwise, it will be put to the next inspection, i.e., contextual analysis. We used regular expressions for fuzzy matching of keyword’s neighbors via some conjunctions, including the prepositions (e.g., about, regarding, response to, relevant to, etc.), verbs (e.g., track, detect, simulate, visualize, etc.) and nouns (e.g., stats, tracker, analysis, model, etc.), and if it matches successfully, we regard it as associated with COVID-19 and thus retain it. 
    After this step, roughly 3,000 repositories are left to check in the final step. 
3) We manually checked the repository to determine whether it should be kept. The first two authors manually analyzed these repositories independently. For the disagreements, a further discussion is performed. In this way, we ended up filtering out 1,190 irrelevant repositories.

\subsubsection{Dataset Overview}
Finally, we have collected 67,079 COVID-19 themed repositories, by the time of July 17, 2020. They are contributed by 67,985 unique contributors in total.

\section{General Overview}
\label{sec:general}

In this section, we present the overview analysis of COVID-19 themed repositories, including the trend of newly created repositories, their activeness, their popularity,
and the distribution of the contributors.

\begin{figure} [t]
\centering
  \includegraphics[width=0.45\textwidth]{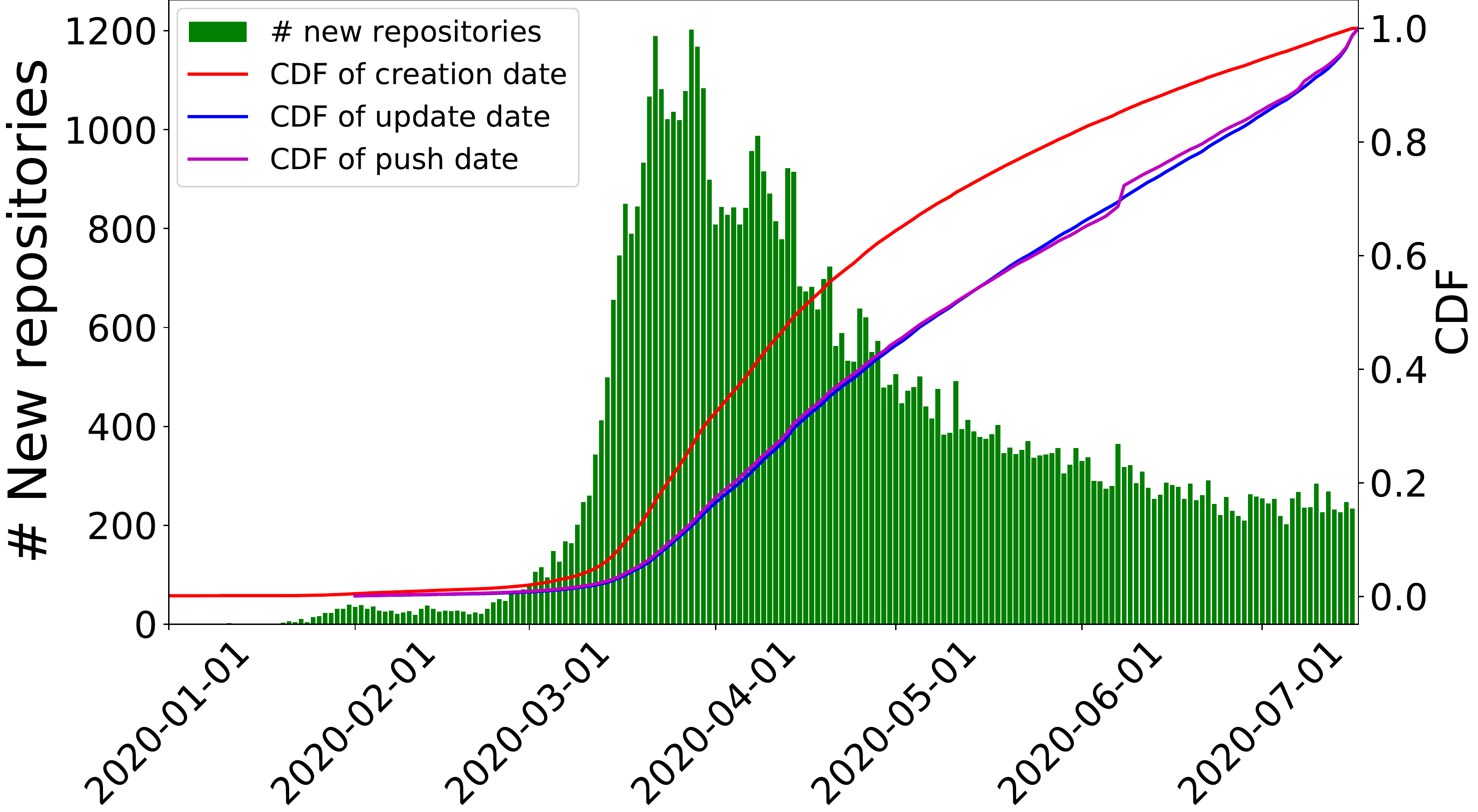}
  \vspace{-0.1in}
\caption{The distribution of creation date, update date and push date.
}
\vspace{-0.2in}
\label{fig:bar-cdf} 
\end{figure}

\subsection{Trend of COVID-19 themed Repositories}

\textbf{Overall Trend.} COVID-19 has rapidly spread around the world and become a global pandemic in early 2020. As it spreads, new repositories regarding the disease emerge in GitHub. We are curious about \textit{when} these repositories were created and \textit{whether} their creation is in line with the trend of the epidemic growth. 
Figure~\ref{fig:bar-cdf} presents the distribution of daily created new repositories, from January to July in 2020. It can be seen that the number of newly created repositories only accounted for less than 0.2\% before mid-January 2020, while the number started to increase sharply from mid-late January and peaked in mid-late March, after which there was a slowly decreasing trend. 
Over 65K (98\%) repositories were created after March 1, 2020, and the day with the most creation was March 28, when 1,202 new repositories were created.
Note that March is also the time when WHO declared COVID-19 a pandemic that started spreading around the world. 

\textbf{Activeness.}
We also pay attention to the activeness of the repositories during the pandemic. 
Through the \emph{updated\_at} and \emph{pushed\_at} indicators which refer to the most recent update and push time of the repository, we summarize the distribution of the latest update date and push date of the repositories, as shown in Figure~\ref{fig:bar-cdf}. 
The distribution of pushed time is almost the same as the updated time (see blue line and purple line in Figure~\ref{fig:bar-cdf}), 
which is expected as the most seen update behavior is push. We thus only consider either of them in our subsequent trend analysis.  
Over 37k (55\%) of the repositories were updated after May 1, 2020, and only 10k (15\%) were updated after July 1, indicating that \textit{a large number of repositories lack continuous activeness.}

\subsection{Repository Popularity} \label{subsec:popularity}

In most previous studies~\cite{borges2015popularity, 7816479}, the popularity of GitHub repositories is usually measured by the number of received stars because the \emph{stargazers} button is an explicit feature for users to manifest their interest or satisfaction with a repository. 
In addition, the number of forks can be seen as a proxy for the importance of a repository in GitHub because \emph{forks} are used to either propose changes to an existing repository or as a starting point for a new repository. 
Besides, we found that many COVID-19 themed repositories were built atop of other repositories, and have shown explicitly links in the \texttt{Readme.md} files, which we called cross-repository reference. 
Thus, our investigation will focus on these three aspects.

\begin{figure} [htbp]
\centering
\subfigure[stars]{
\label{fig:stars}
  \includegraphics[width=0.23\textwidth]{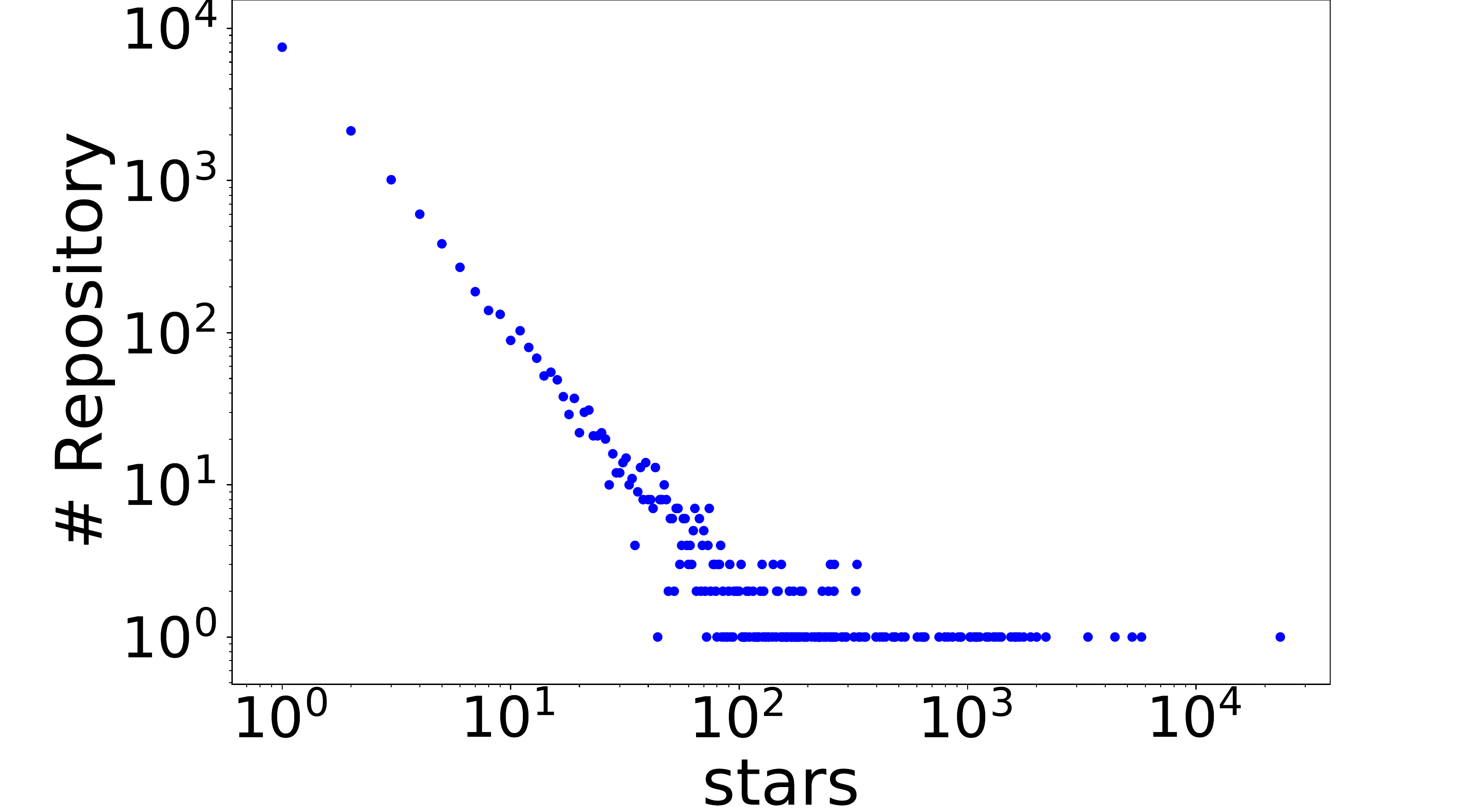}}
\subfigure[forks]{
\label{fig:forks}
  \includegraphics[width=0.23\textwidth]{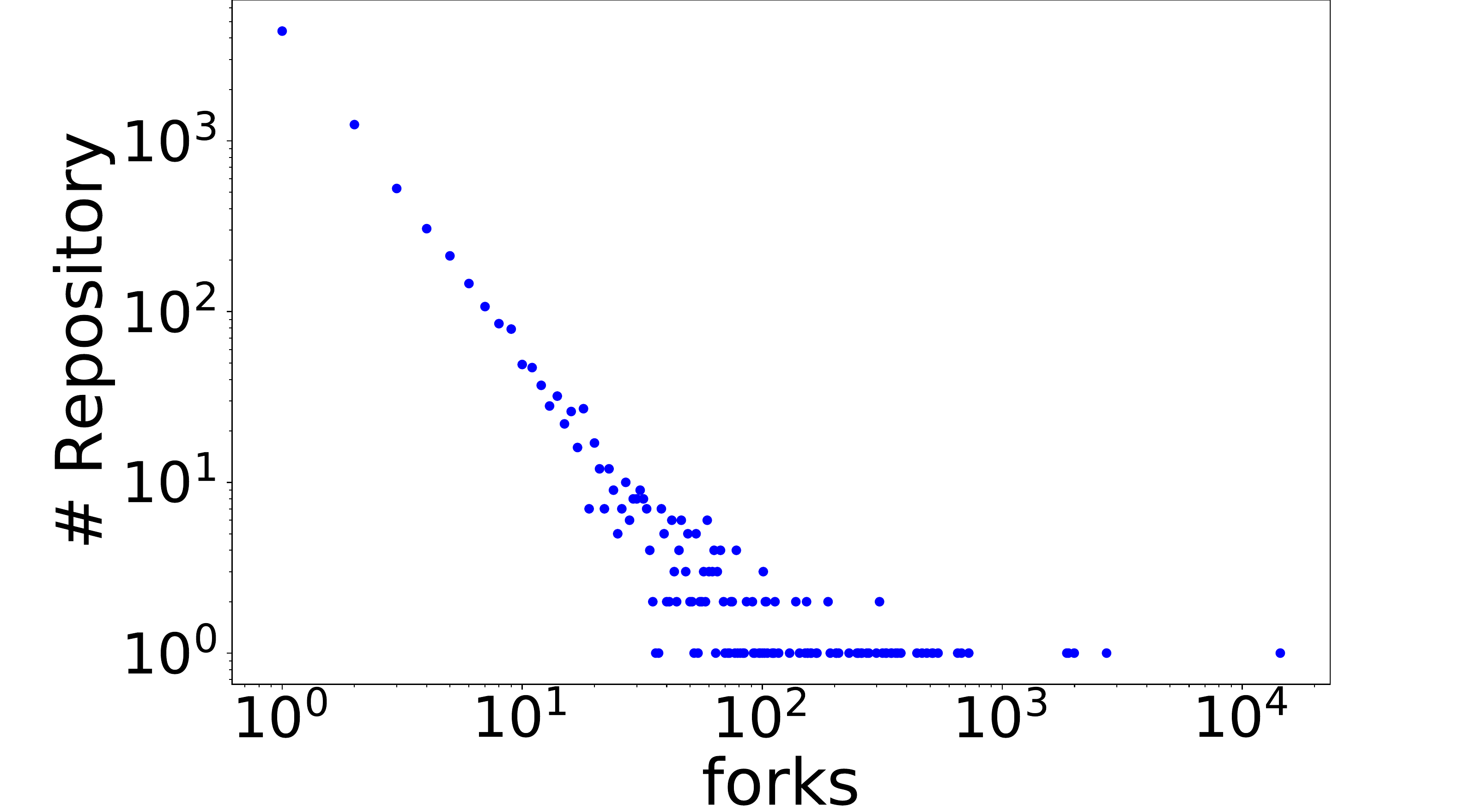}}  
  \vspace{-0.1in}
\caption{The distribution of repository popularity.}
  \vspace{-0.1in}
\label{fig:star-fork} 
\end{figure}

\textbf{The number of received stars.}
We counted the number of repositories per star and found that the 0-star accounted for the vast majority, up to 79.6\%, followed by the 1-star with 11.2\%. Figure~\ref{fig:stars} shows the distribution of the number of repositories with more than one star.
Obviously, it follows the typical \textit{power law distribution}, i.e., there are only a few repositories with large number of stars, while most repositories receive few stars. The most representative repository, CSSEGISandData/COVID-19\footnote{https://github.com/CSSEGISandData/COVID-19}, which is a COVID-19 data repository hosted by the Center for Systems Science and Engineering (CSSE) at Johns Hopkins University, 
has received 23,352 stars by the time of our study.

\textbf{The number of forks.}
The forks have almost the same power law distribution with the stars, which is consistent with our intuition (see Figure~\ref{fig:forks}). There is a high percentage of 0-fork repositories (88.5\%), and the 0-fork and 1-fork repositories make up the vast majority (95\%). 
The repository with the most forks is exactly the repository with the most stars, i.e., the SSEGISandData/COVID-19, which has been forked 14,392 times as of this writing.

\begin{figure} [t]
\centering
  \includegraphics[width=0.6\textwidth]{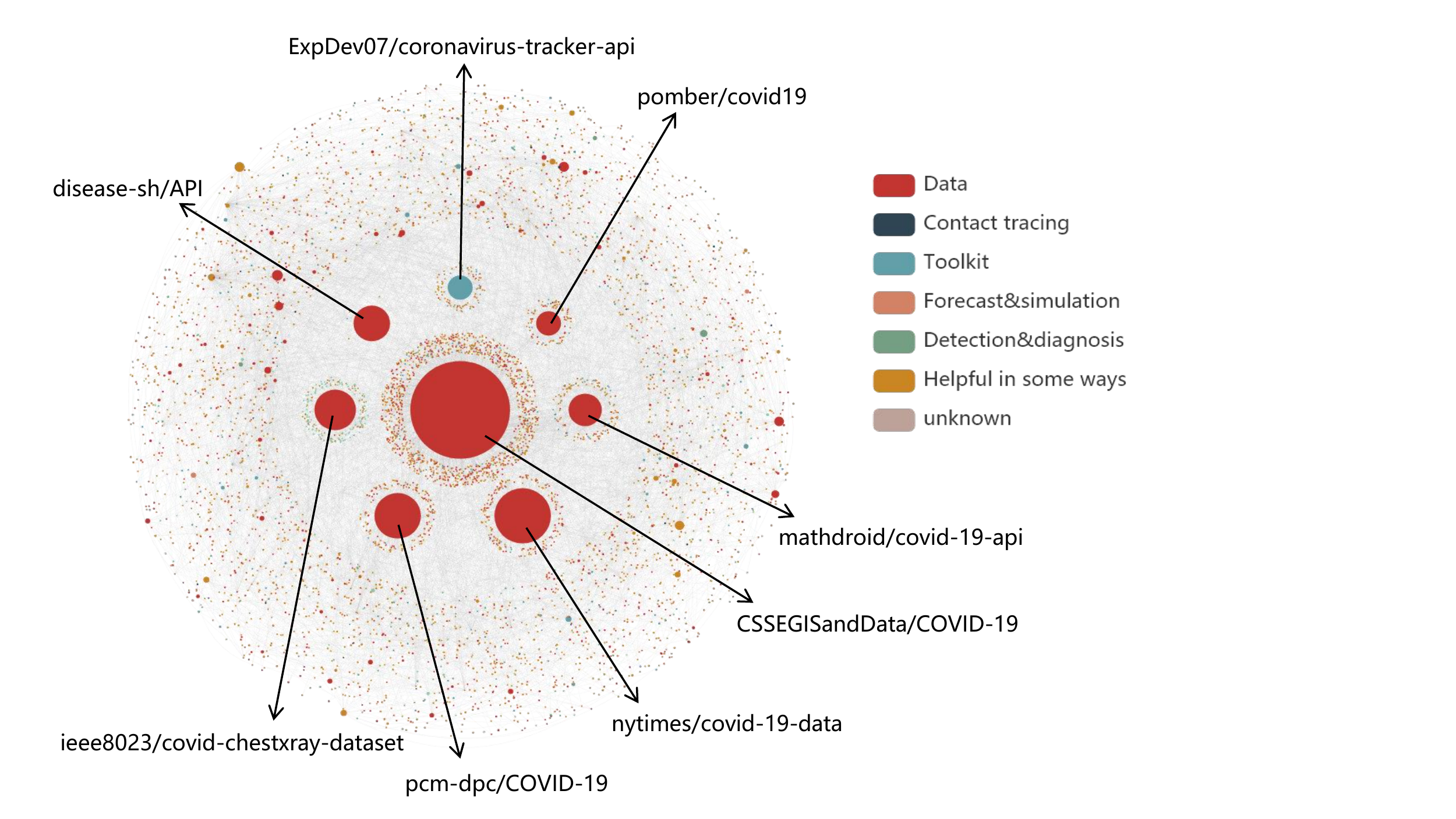}
  \vspace{-0.15in}
\caption{Cross-referencing of COVID-19 repositories.}
\vspace{-0.2in}
\label{fig:link} 
\end{figure}

\textbf{Cross-repository Reference.}
In addition to the descriptions, we have collected the \emph{Readme.md} files of all the repositories in our dataset. We extracted the links from these files and tried to investigate referencing relationships among the COVID-19 themed repositories. It turns out that, 4,811 repositories contain links in their \emph{Readme} to other repositories in our dataset, and 2,224 repositories have been referenced by others. 
Figure~\ref{fig:link} presents the cross-reference relations among the repositories. The larger the node, the more the corresponding repository is referenced.
Note that we have classified the repositories into six categories (to be detailed in Section~\ref{sec:classification}), and 
the color of a node represents the category of the corresponding repository. 
As can be seen in Figure~\ref{fig:link}, the top 7 most referenced repositories are of the data category. 
The most frequently referenced repository is again the CSSEGISandData/COVID-19, which has been referenced by 1,640 repositories.

\subsection{The Distribution of Contributors}
\begin{figure} [t]
\centering
\vspace{-0.1in}
  \includegraphics[width=0.4\textwidth]{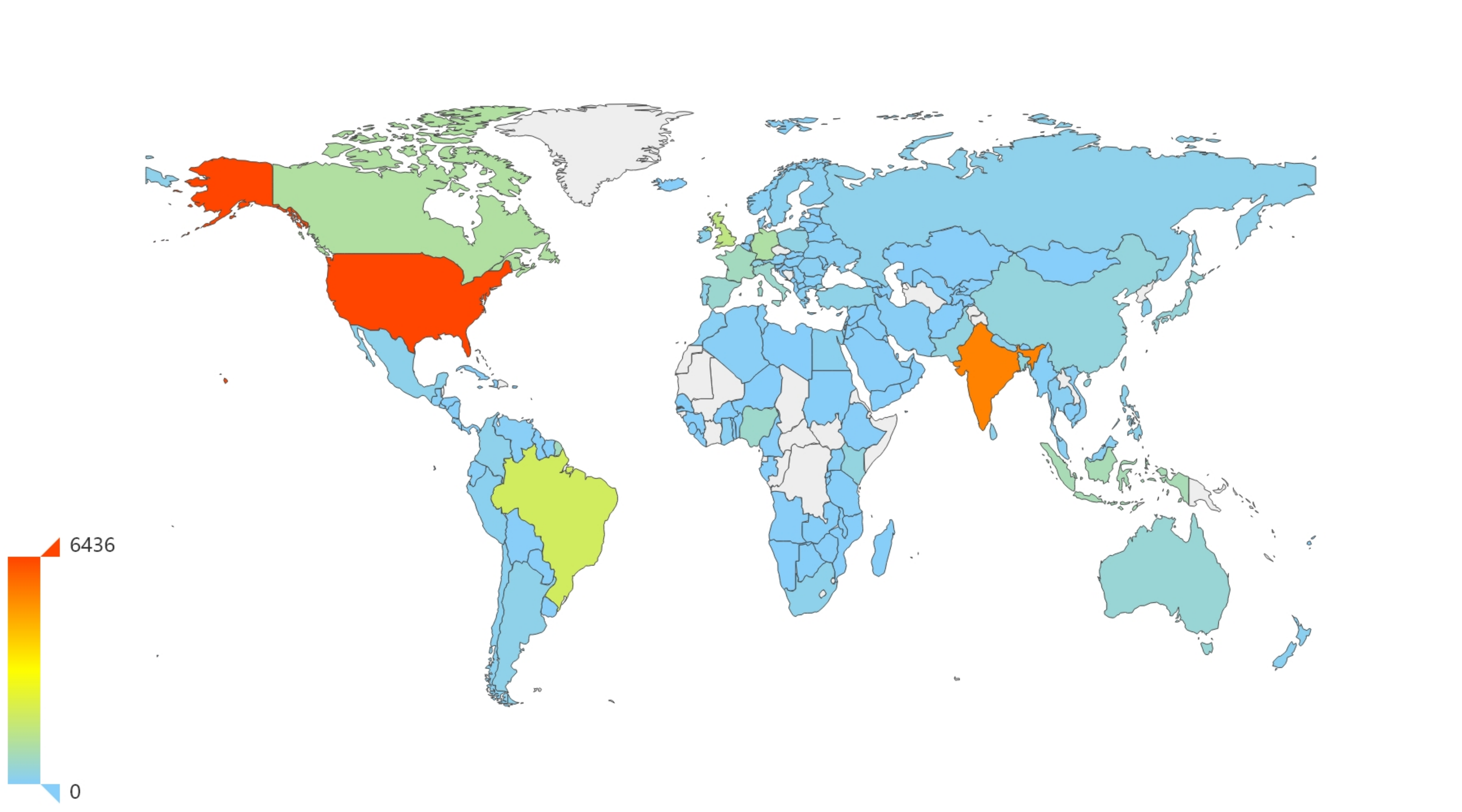}
  \vspace{-0.2in}
\caption{The country distribution of contributors}
\vspace{-0.2in}
\label{fig:contributor_map} 
\end{figure}

Next, we investigate the distribution of the 67,985 collected contributors.
As aforementioned, we gathered their location information during our data collection, while unfortunately 35,537 (52\%) of them did not provide their location data. 
Among the location data provided, there are some invalid ones like ``somewhere'', ``universe'', ``planet'', ``internet'', ``home'', etc. 
Besides, different formats/representations of the location are provided, including city name only, country name only, and both. 
Thus we started with filtering invalid location names, and then used Geotext~\cite{geotext} to convert all the valid locations into country names. 
Finally we acquired a total of 31,296 valid locations mapping over 200 countries. 
A map displaying the country distribution of these contributors is shown in Figure~\ref{fig:contributor_map}. The top 10 countries in regards to the number of contributors are the United States (6,436), India (5,389), Brazil (1,992), the United Kingdom (1,529), Canada (1,132), Germany (1,089), Indonesia (877), France (764), Nigeria (606) and Italy (581), respectively. \textit{The distribution of contributors is inline with the distribution of COVID-19 cases around the world}, i.e., United States, India and Brazil are top countries with regard to the confirmed COVID-19 cases.

\begin{figure*} [htbp]
\centering
\subfigure[US]{
\label{fig:us}
  \includegraphics[width=0.32\textwidth]{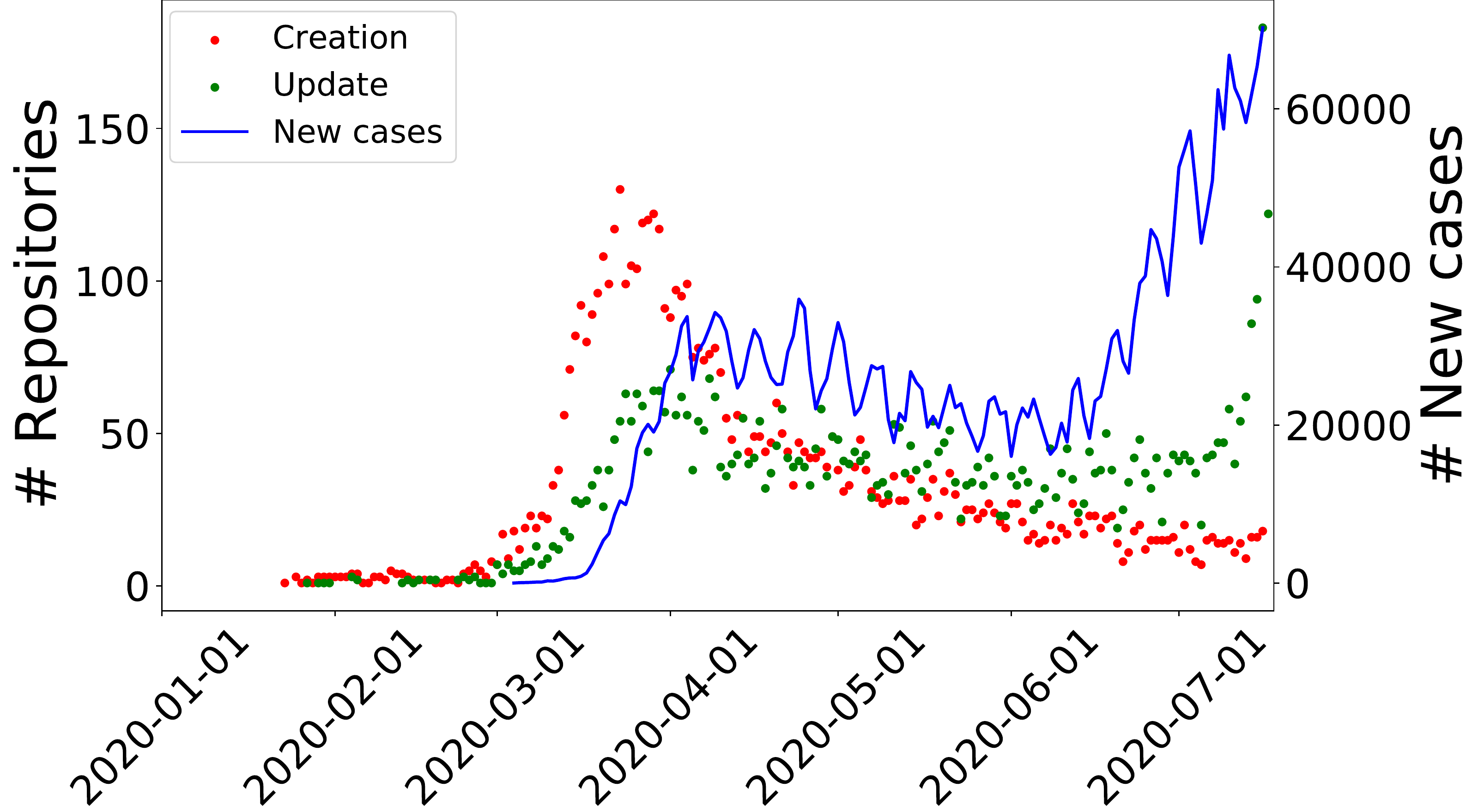}}
\subfigure[India]{
\label{fig:indai}
  \includegraphics[width=0.32\textwidth]{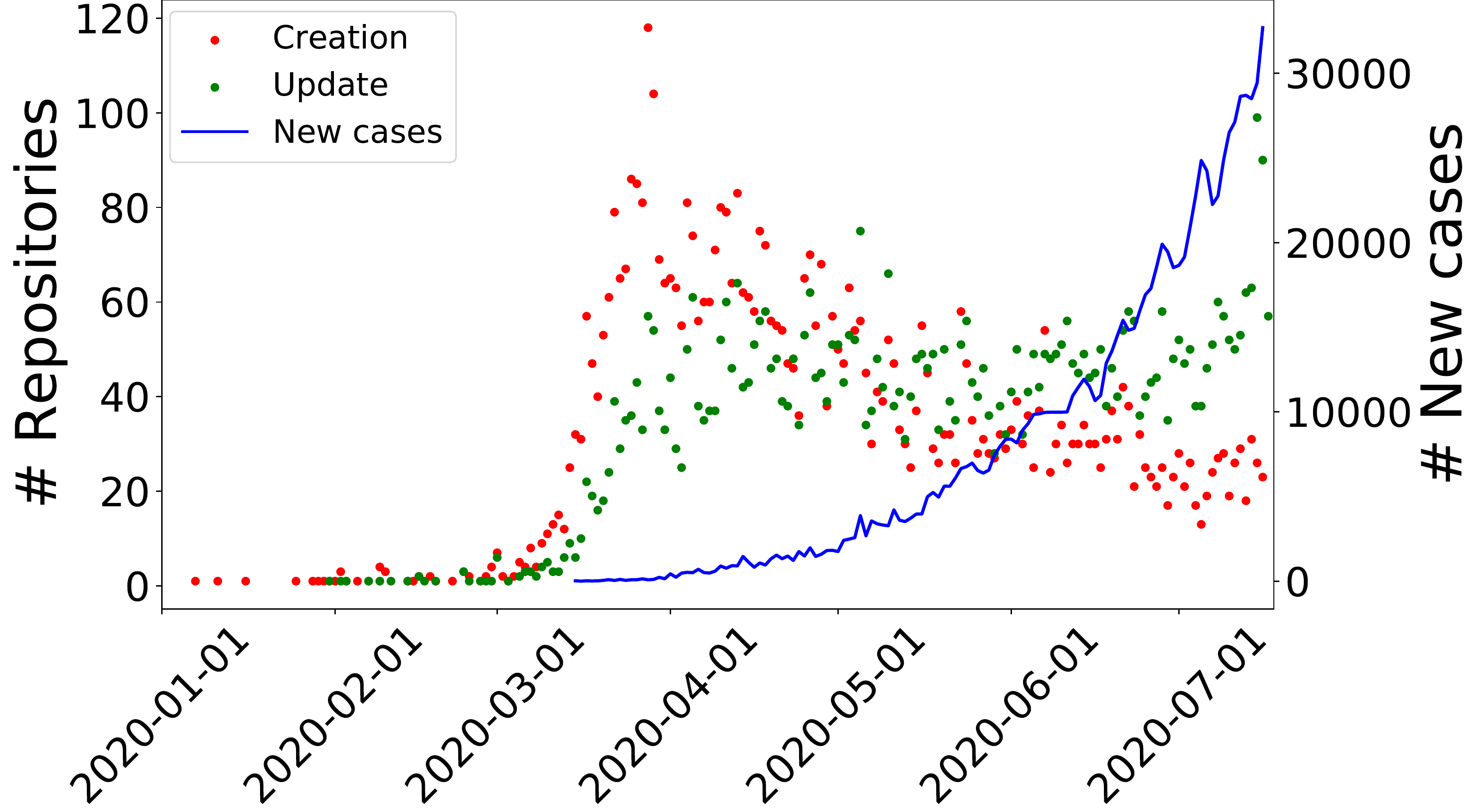}}  
\subfigure[Brazil]{
\label{fig:brazil}
  \includegraphics[width=0.32\textwidth]{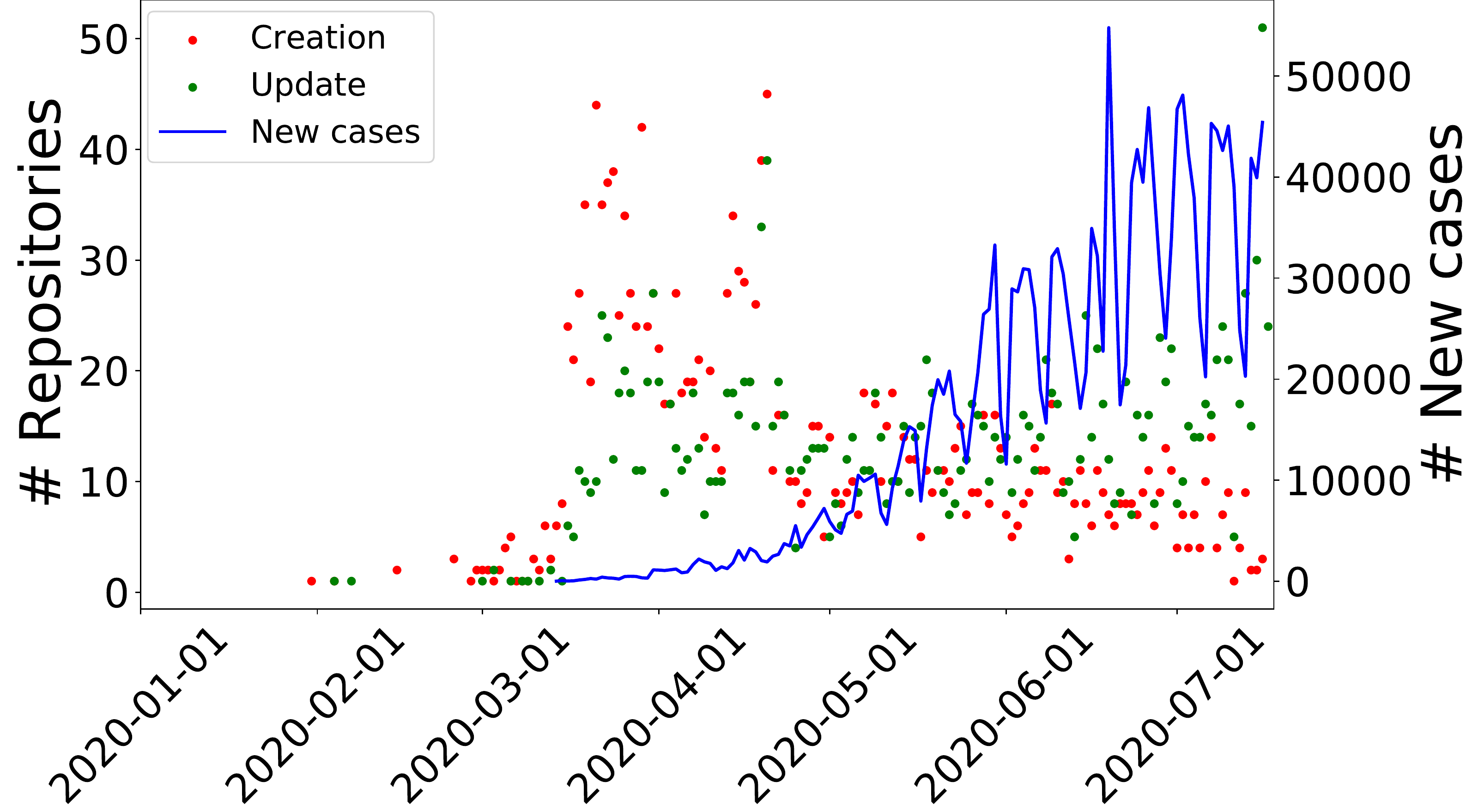}}  
\subfigure[UK]{
\label{fig:uk}
  \includegraphics[width=0.32\textwidth]{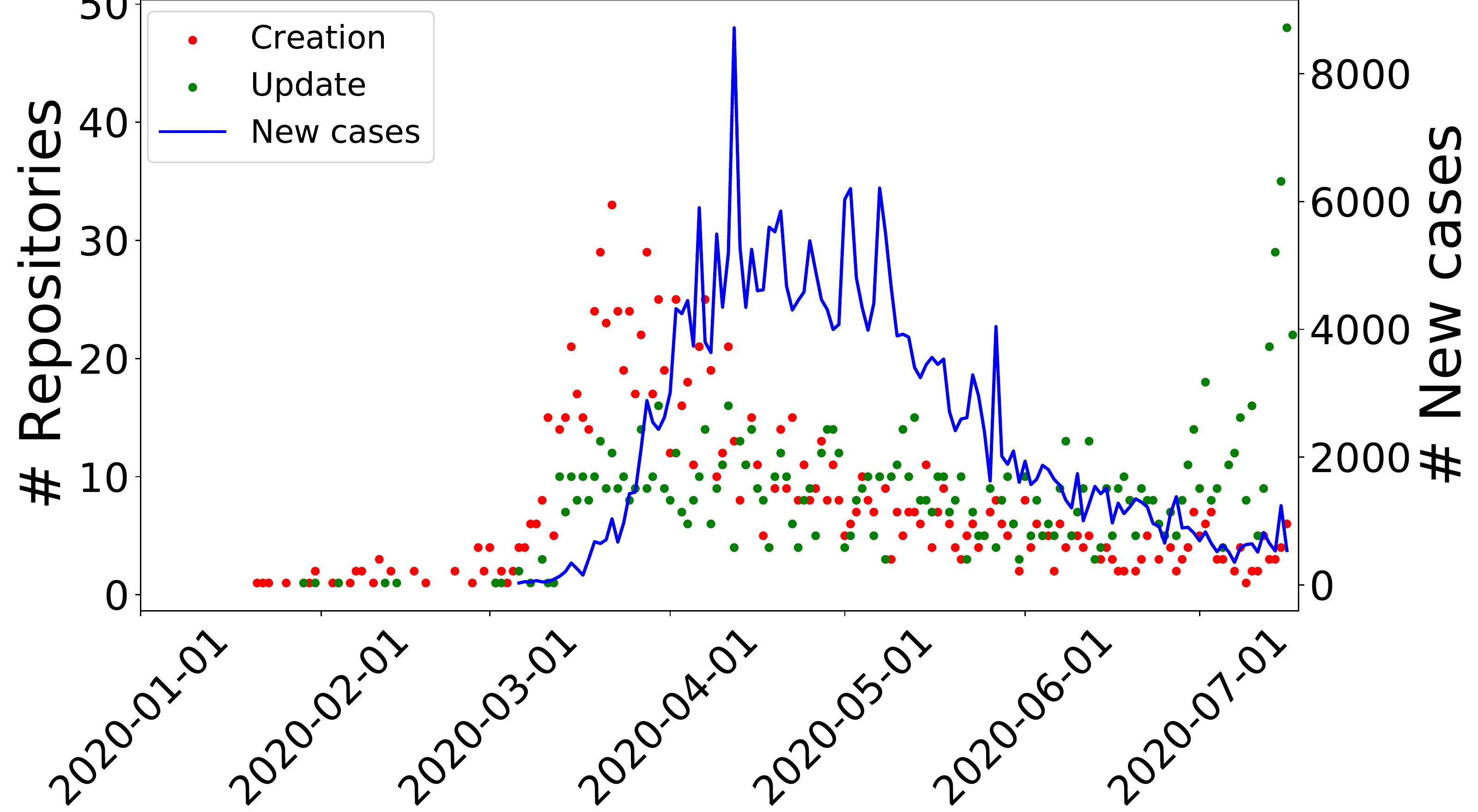}}  
\subfigure[Canada]{
\label{fig:canada}
  \includegraphics[width=0.32\textwidth]{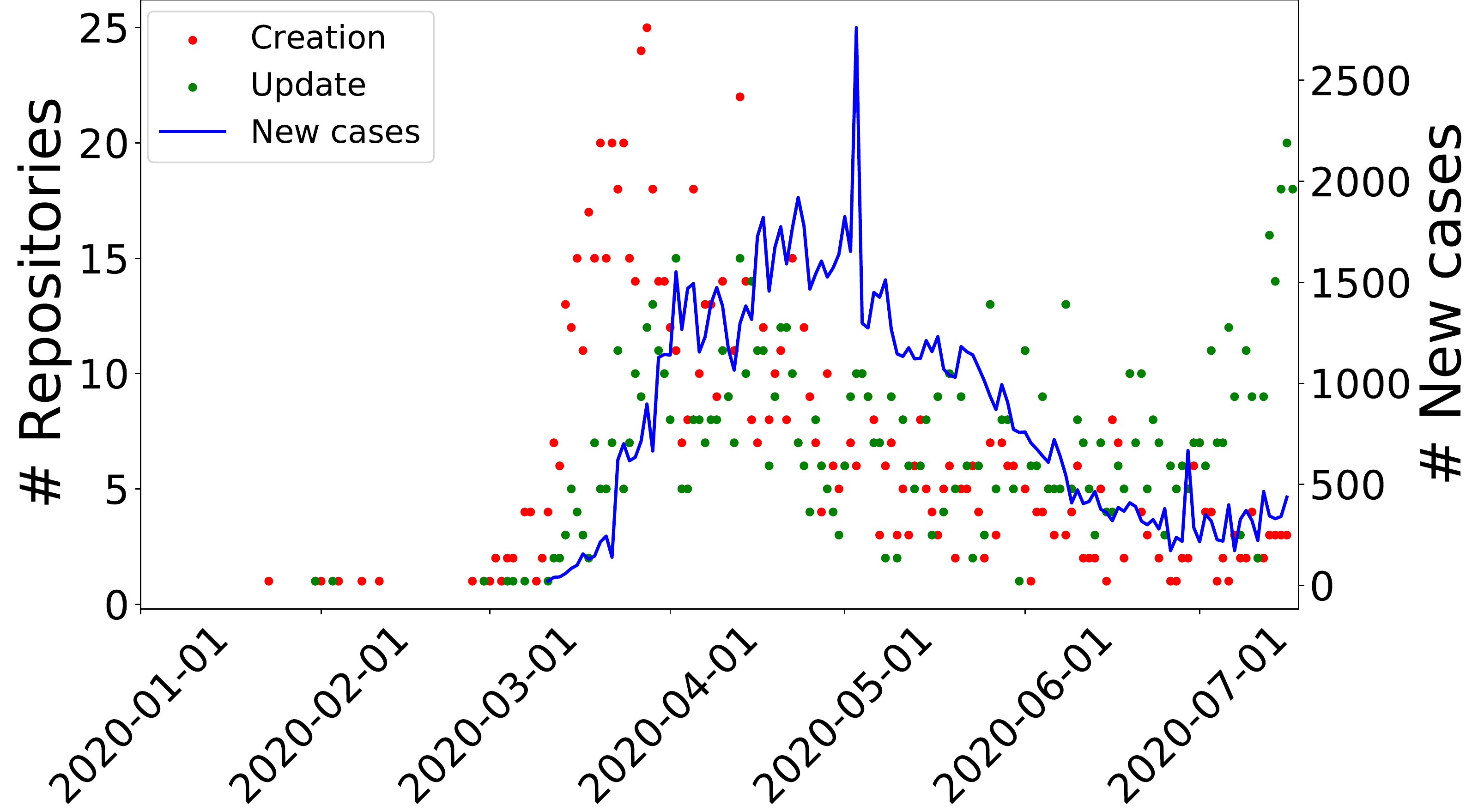}}     
\subfigure[Germany]{
\label{fig:germany}
  \includegraphics[width=0.32\textwidth]{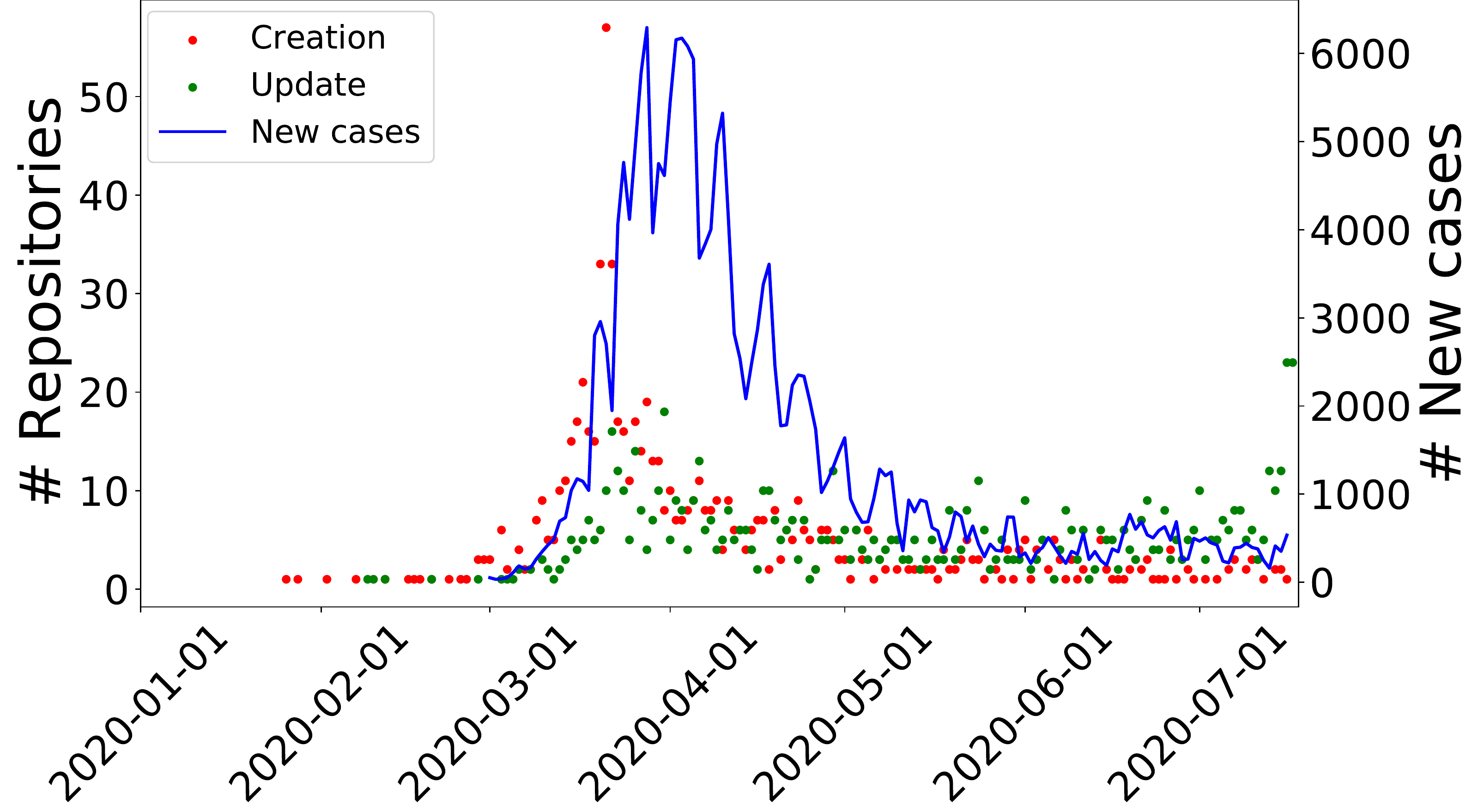}}
  \vspace{-0.1in}
\caption{The number of newly created repositories and newly confirmed cases over time in each country}
\vspace{-0.1in}
\label{fig:country-comparison} 
\end{figure*}

\subsection{Per Country Analysis}
Considering that the emerging time of the COVID-19 pandemic varies from country to country, it is thus valuable to investigate \textit{whether the number of daily new created repositories matches the number of daily new confirmed cases}. On account of this, we conducted case studies in the top 6 countries with the largest number of contributors, i.e., the United States, India, Brazil, the United Kingdom, Canada and Germany. We define which country a repository belongs to by its contributors' locations. If all the contributors (one or more) of a repository are located in the same country, we treat the repository as belonging to that country; if the different contributors of a repository are located in multiple countries, we take the repository as \textit{transnational}. 
In this way, we worked out the results as shown in Table~\ref{tab:contributors}. 
Obviously, the distribution of repositories is in line with the distribution of contributors across countries. We notice that \textit{most of the repositories are contributed by developers of a sole country, while only 1,490 of them are transnational repositories that contributed by developers from multiple countries}.

We next investigate the correlation between the development process and the COVID-19 evolvement in each country. 
We collected the number of new infections per day as complete as possible for each country. We show the comparison chart of 6 countries in Figure~\ref{fig:country-comparison}. In each plot, the red scatter shows the number of newly created repositories each day and the blue line presents the number of newly confirmed cases per day. 
Beyond this, we also take the latest updated time into account and plot it with green scatter. 
From the graphs, we can derive at least two interesting findings: 
1) Regardless of countries, the trend is relatively similar in both the number of newly created repositories and latest update repositories, and even consistent with their overall trends~(see Figure~\ref{fig:bar-cdf}). 
This partly reflects a quick reaction to the public health emergency in these countries. 
2) Taking the blue line into account, we can find that in some countries, the line and scatters show a strong consistency in the trend (overall or in stages). For four countries, i.e., US, UK, Canada and Germany, we can observe that the blue line, red scatter and green scatter have \textit{significant up-trends that almost overlap and are followed by a slow decline}. While for India and Brazil, we can see a lot of attention has been paid to the outbreak when it became a pandemic in March, even though there was not yet a widespread outbreak in their countries at the time. \textit{This reflects the fact that national responses are not always fully aligned with the epidemic situation in their own countries. Outbreaks in other countries can serve as a cautionary tale for a country.}

\begin{table}[t]
\centering
\small
\caption{Distribution of contributors and repositories in top 6 countries.}
\vspace{-0.1in}
\resizebox{0.49\textwidth}{!}{
\begin{tabular}{lccccccc}
\toprule
~ & US & India &  Brazil &  UK & Canada & Germany & transnational \\ 
\midrule
Contributors & 6,436 & 5,389 & 1,992 & 1,529 & 1,132 & 1,089 & - \\
Repositories & 5,828 & 5,775 & 1,860 & 1,286 & 997 & 815 & 1,490 \\
\bottomrule
\end{tabular}
}
\vspace{-0.15in}
\label{tab:contributors}
\end{table}

\begin{framed}
\noindent \textbf{Answer to RQ1:} 
\textit{
There are over 67K COVID-19 related repositories in GitHub by the time of our study, and 98\% of them were created after March 1, the time when the coronavirus becomes a pandemic. As the outbreak spread extensively, a significant increase in the creation and alteration of related repositories in GitHub instantly followed, indicating the open source community's rapid response to the pandemic.
}
\end{framed}

\section{Classifying COVID-19 Repositories} \label{sec:classification}

In this section, we seek to understand the focus of collected COVID-19 repositories. 
We first rely on manually efforts to create a taxonomy.
Then, we adopt an automated classification method to classify them.

\subsection{Taxonomy}

We conduct a manual investigation to understand the subjects and contents of the COVID-19 themed repositories in order to create the taxonomy. 
As it is infeasible for us to inspect all collected repositories, we seek 200 representative ones for our manual analysis. 
As we discussed in Section~\ref{subsec:popularity}, the number of stars partially indicates the popularity and quality of a repository, so we picked the top 200 stared ones. 
Through extensive manual analysis, we create a taxonomy including the following six categories: 

\begin{itemize}
    \item[C1] \textbf{Data}. A large portion of the 200 samples are data-related, including COVID-19 cases statistics, image datasets, data visualization and analysis, etc. 
    \item[C2] \textbf{Contact tracing}. 
    Contact tracing is a key technique used by public health authorities to contact and give guidance to anyone who may have been exposed to COVID-19 cases. Some repositories in our collection are open-source contact tracing apps or frameworks. For example, the repository google/exposure-notifications-server\footnote{https://github.com/google/exposure-notifications-server} implements the Exposure Notifications API and provides reference code for working with Android and iOS apps.
    \item[C3] \textbf{Toolkit.} 
    A number of tracking toolkits for COVID-19 epidemic have emerged in GitHub, including mobile apps, APIs, crawlers, Python packages, etc. 
    For example, some repositories
    provide APIs for tracking the outbreak around the world or in  specific regions; 
    some repositories release tracker apps with real-time information about the novel coronavirus.
    \item[C4] \textbf{Forecast and simulation.}
    To assist in developing coping strategies and decision making,
    there are a number of open source repositories dedicated to forecasting, estimating or simulating the disease spread. 
    \item[C5] \textbf{Detection and diagnosis.} In response to the tremendous global toll of COVID-19, researchers have rapidly mobilized to investigate solutions for detection, diagnosis, and treatment. 
    For example, some repositories
    propose to diagnose COVID-19 from chest CT or X-Rays, and some of them 
    implement web services that allow users to take an at-home COVID-19 antibody test.
    \item[C6] \textbf{Helpful in some ways}. 
    Repositories in this category may not directly target COVID-19, but have a positive effect on people's life in some ways through the difficult period including education, travel, medical equipment, business, entertainment, communication, etc.

\end{itemize}

\vspace{-0.1in}
\begin{table}[t]
\centering
\small
\caption{A taxonomy of COVID-19 themed repositories.
}
\vspace{-0.1in}
\resizebox{0.49\textwidth}{!}{
\begin{tabular}{lcc}
\toprule
Category & Keywords \\ 
\midrule
Data & data,stat,case,dataset,visual,analysi,updat, daili, death \\
Contact tracing & trace, contact, exposur, notif, ble, bluetooth, privaci, infect  \\
Toolkit & track, tracker, command, built, line, global, latest, statu  \\
Forecast \& simulation & simul, model, forecast, predict, spread, estim, seir, trend  \\
Detection \& diagnosis & x-ray, chest, detect, imag, ct, neural, scan, diagnosi, lung \\
Helpful in some ways & help, ventil, deliveri, shop, resourc, busi, support, suppli \\
\bottomrule
\end{tabular}
}
\vspace{-0.1in}
\label{tab:category-keywords}
\end{table}

\subsection{Classification Approach}

We next seek to automatically classify the repositories. 

\textbf{Ground truth.}
We first built a ground truth dataset to be used for training our classification model. 
It is intended to include 600 labelled repositories, 100 per category. 
We built it by manually labeling collected repositories one by one in descending order of the number of stars, till each category has been assigned 100 repositories. 
Considering there are multiple languages in the descriptions, we used Google Translation to translate all the descriptions into English before labelling.
The labeling was done by the first two authors independently too.
A discussion was performed to reach a consensus on the labels in the end.
After obtaining the ground truth, we utilize the labeled descriptions to build a classification model that enables us to classify other repositories.

\textbf{Approach Overview.}
We adopted a distance-based classification methodology.
Below we brief each step in our approach. 

1) \textit{Keyword generation and feature set extraction}. 
The first step of our approach is to generate keywords for each category such that the characteristics of each category can be represented by its keywords. 
Considering that most of the descriptions have few words, we merged all the 100 descriptions in a category into a single text, resulting in six documents as our training data. 
Each of them has its category name as its label. 
We extended emojis and those general-thus-indistinguishing words like "COVID-19" and "coronavirus" into the stopwords list (downloaded form www.nltk.org), so that they are removed from the training data prior to our keyword generation. 
All terms in the training data are then taken to calculate their weights in differentiating the six documents
using the TF-IDF~\cite{7007894},
which is a widely used method to find out how important a word is to differentiate the labelled documents from each other.
We then took the top 20 keywords for each category and merged them into a feature set of length 117 (after deduplication). 
We list several representative keywords in Table~\ref{tab:category-keywords}.

2) \textit{Classification}. 
The similarity of documents can be expressed by the angle or distance between their feature vectors - the smaller the angle or distance means the more similar the two documents~\cite{Artama_2020}. 
Thus, we combine TF-IDF and the cosine similarity
for classification. 
Given a repository to classify, we take its description along with the training documents as input to the TF-IDF method, which generates vectors including the weights of all terms that appear in all documents. 
These vectors are projected to our feature set (of length 117), so that the feature vectors of all six categories and the tested repository are derived. 
We then measure their cosine distance, and the tested repository is classified into the nearest category. 

3) \textit{Cross validation}. 
We adopted a ten-fold cross validation to measure its effectiveness. 
As shown in Table~\ref{tab:validation-results}, our method achieves a promising performance, with an overall accuracy of 92\%.
It is thus feasible to apply our classification model to large scale repository classification.

\vspace{-0.1in}
\begin{table}[t]
\centering
\small
\caption{The precision and recall of the cross validation.}
\vspace{-0.1in}
\begin{tabular}{l|cccccc}
\toprule
Category & C1 & C2 & C3 & C4 & C5 & C6 \\ 
\midrule
Precision & 0.83 & 0.97 & 0.96 & 0.94 & 0.95 & 0.94 \\
Recall &  0.92 & 0.94 & 0.95 & 0.91 & 0.88 & 0.92 \\
\bottomrule
\end{tabular}
\vspace{-0.1in}
\label{tab:validation-results}
\end{table}

\subsection{Overall Classification Results}

We applied our classification method to the rest of the dataset. 
We only focus on the six categories we have identified, as they are the most dominant and representative (recall that we identified them based on the top 200 stared repositories).  
There are also a number of categories whose description is either very short (even without description), unclear or uncharacteristic such that it is infeasible to classify them into any of our categories. 
Therefore, we conducted a filtering on our dataset based on the 117 keywords we have obtained. 
We consider introducing a threshold \emph{n}, and filer out the repositories whose description contains fewer than \emph{n} keywords. 
As such, the key becomes to figure out an appropriate \emph{n}. 

To this end, we analyzed the 600 labeled ground truth. 
We learned that the minimum number of keywords included in a description is 1 (with 130 descriptions), the maximum is  17 (with 2 descriptions), and the average is 3.08. 
We also noticed that the descriptions with 2 keywords are the most common (with 161 descriptions). 
It turns out that 21.6\% of the repositories are possible to be classified by a single keyword, so we set \emph{n} to 1. 
However, not every keyword has the same capacity to distinguish categories, so we grouped the keywords into two levels, i.e., primary and secondary.
The primary keyword means that the keyword itself could be relied on to categorize the repository. 
We identified them from those descriptions containing only one keyword. Overall, we obtained 34 such keywords, including "stat", "data", "visual", "track", "forecast", "predict", "detect", etc.;  
the remaining 83 are taken as secondary keywords. 
With obtained keywords, we filtered the dataset in such a way that, if the description contains at least one primary keyword or at least two secondary keywords, we keep it. 
Through this, we obtained 19,739 classifiable repositories. 
We had a sampling to test our approach, and found that only 5 out of 200 randomly selected repositories cannot be classified into any of our six categories, which is acceptable. 
Finally, we have these repositories categorized using our classification method. 
The result is shown in Table~\ref{tab:classification-results}.
Obviously, data category dominates the repositories (55\%). Only 276 repositories are related to contact tracing 
techniques.

\begin{table}[t]
\centering
\small
\caption{Overall Classification Result.}
\vspace{-0.1in}
\begin{tabular}{l|cccccc}
\toprule
Category & C1 & C2 & C3 & C4 & C5 & C6\\ 
\midrule
\# repository & 10,882 & 276 & 3,276 & 2,231 & 1,064 & 2,010 \\
\bottomrule
\end{tabular}
\vspace{-0.1in}
\label{tab:classification-results}
\end{table}

\begin{figure} [t]
\centering
\subfigure[Create\_at]{
\label{fig:category-create}
  \includegraphics[width=0.23\textwidth]{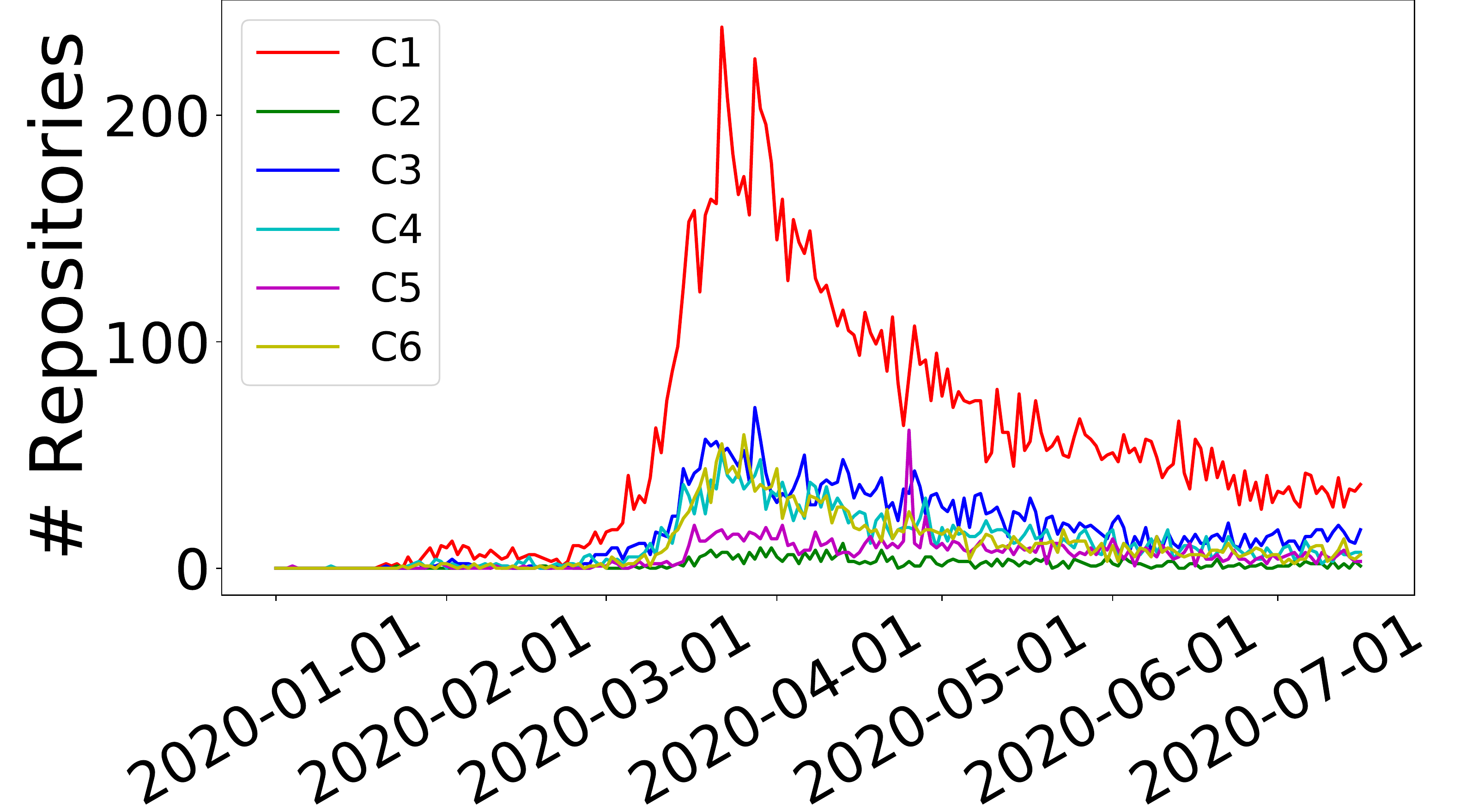}}
\subfigure[Commit]{
\label{fig:category-commit}
  \includegraphics[width=0.23\textwidth]{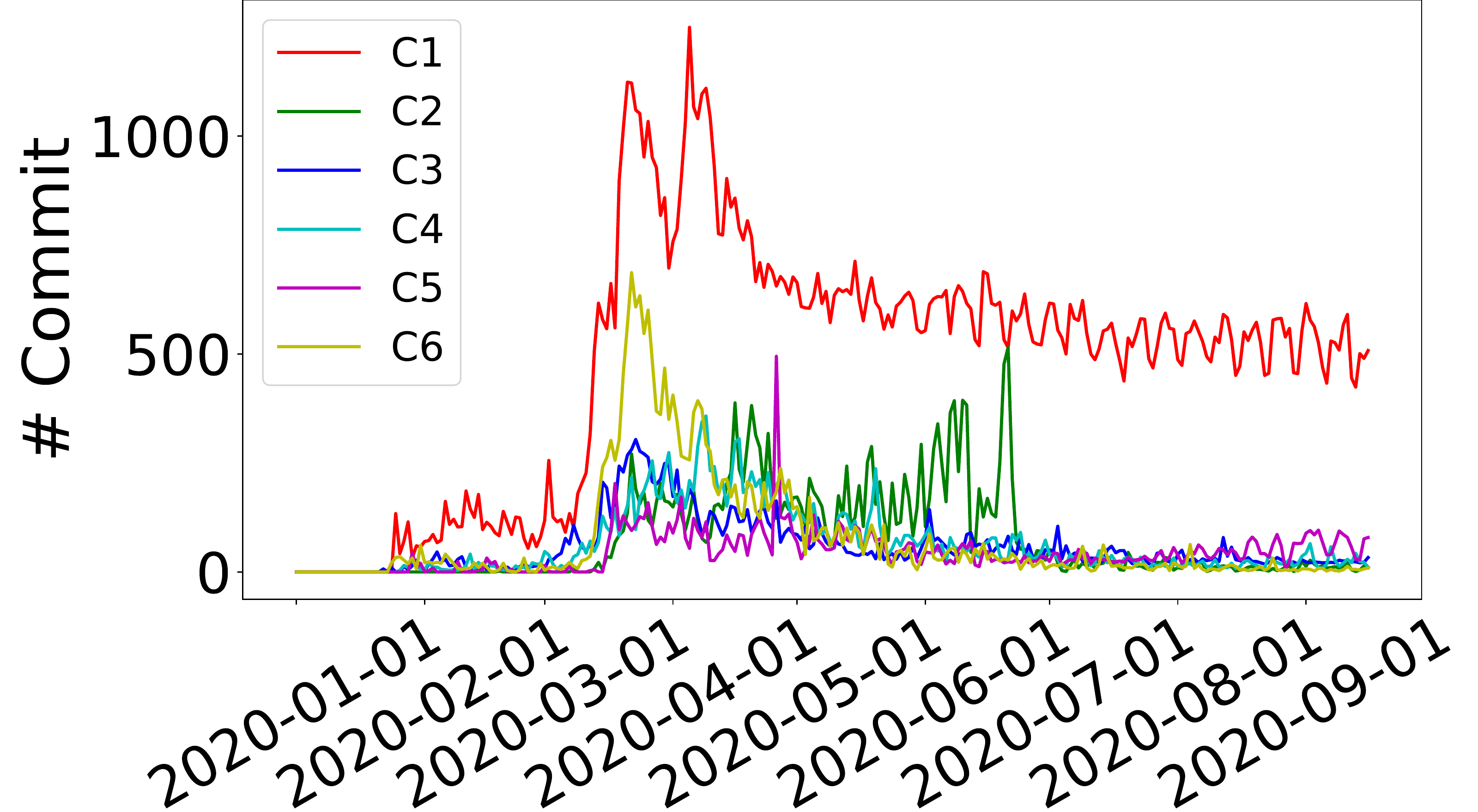}}  
  \vspace{-0.1in}
\caption{The distribution of creation date and \# of commits in each category}
\vspace{-0.2in}
\label{fig:category-analysis} 
\end{figure}

\subsection{Category-level analysis}
We next take a deeper dive into each category. 
First, we explore \textit{whether there are differences in the timing of the prevalence of different categories of repositories.}  Figure~\ref{fig:category-create} shows the distribution of creation date for all the repositories in each category separately. 
Most categories have similar trends and are consistent with the overall trends of creation time (see Figure~\ref{fig:bar-cdf}), 
while there is a distinct difference for the ``Detection and diagnosis'' category, where the line peaks at the end of April. 
\textit{This indicates that the repositories involving detection and diagnosis elements flourished slightly later than other categories, which is reasonable}.

\begin{figure*} [t]
\centering
\subfigure[Overall]{
\label{fig:overall-contributor}
  \includegraphics[width=0.3\textwidth]{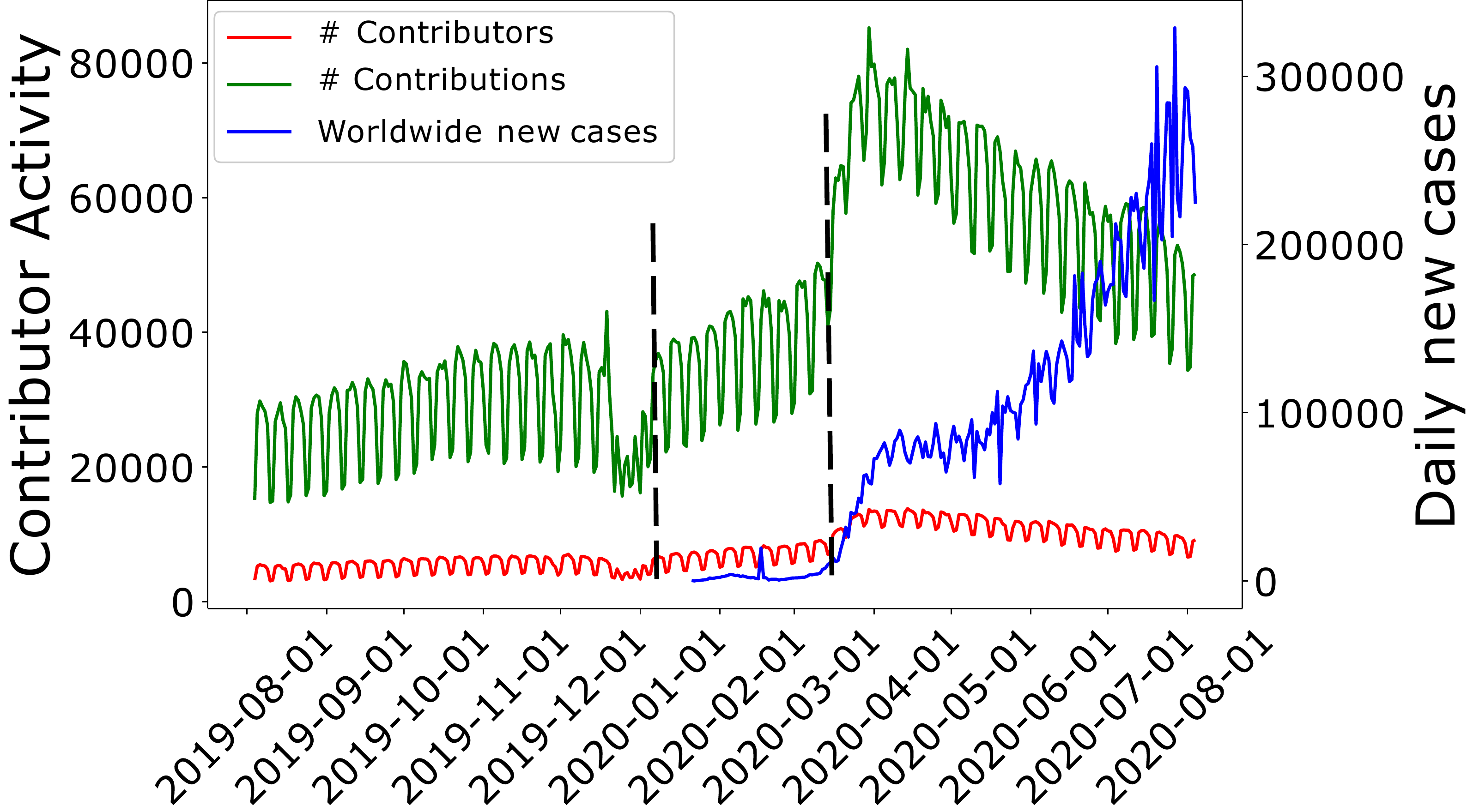}}
\subfigure[US]{
\label{fig:us-contributor}
  \includegraphics[width=0.3\textwidth]{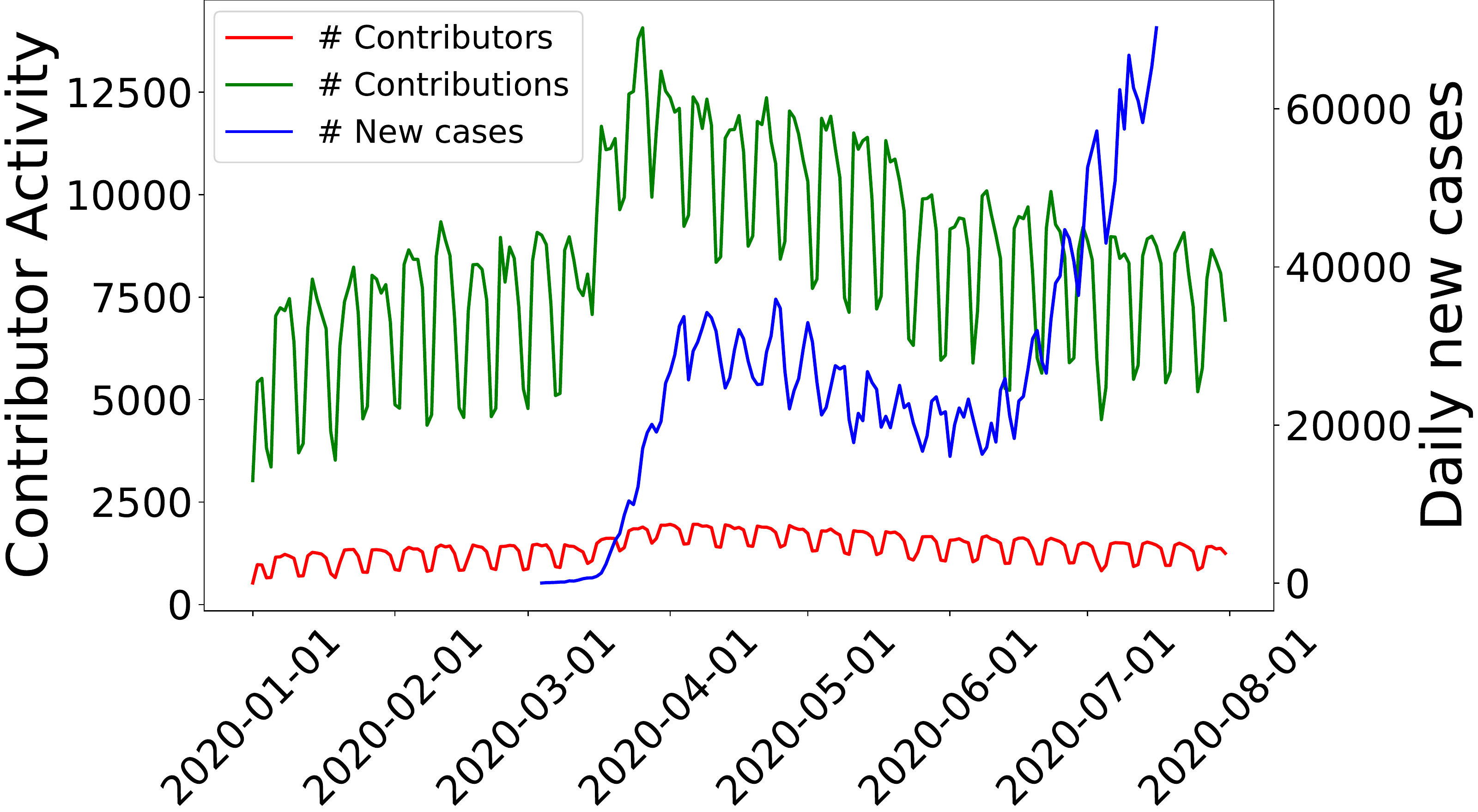}}  
\subfigure[UK]{
\label{fig:uk-contributor}
  \includegraphics[width=0.3\textwidth]{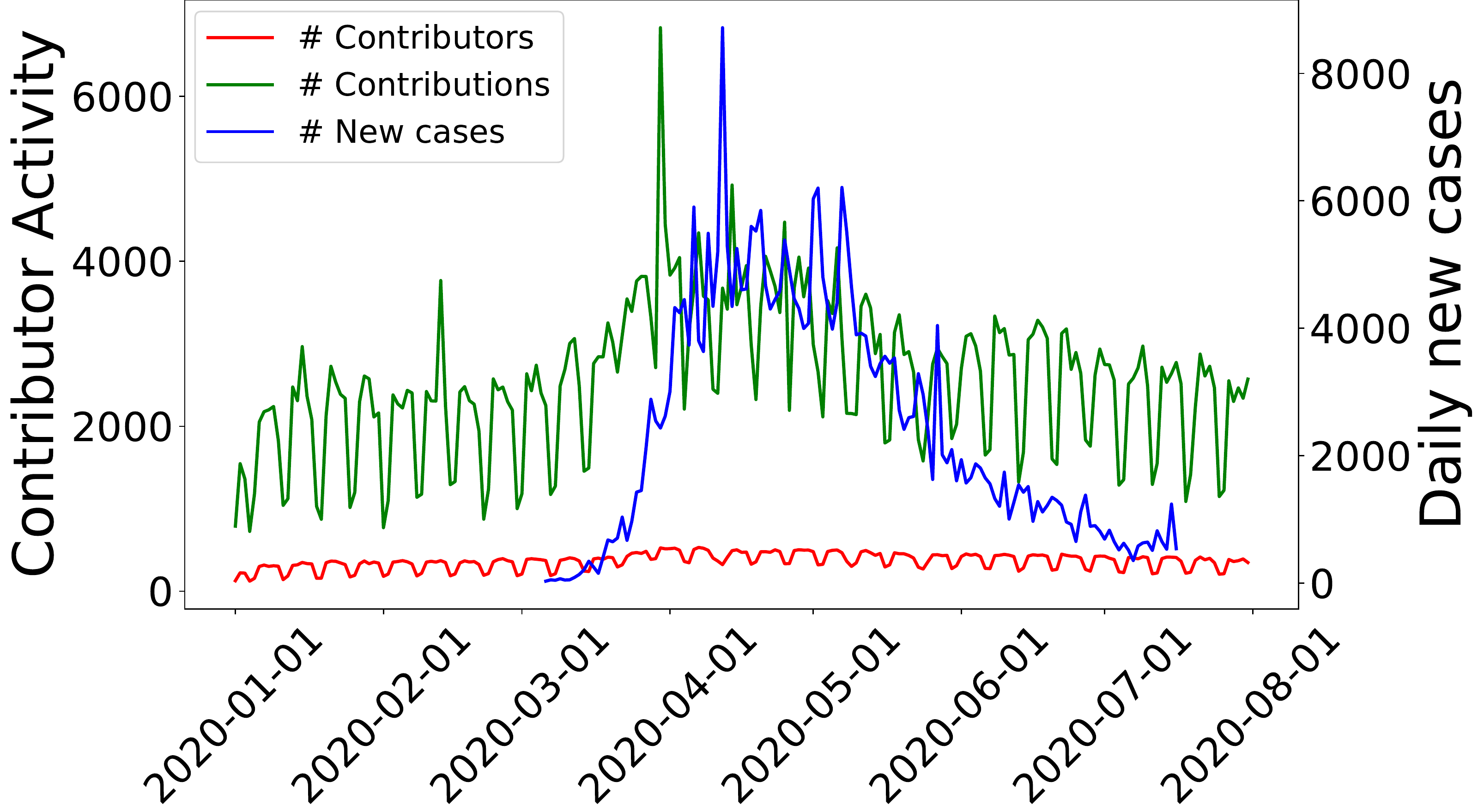}}  
  \vspace{-0.1in}
 \caption{The activity of contributors over the time}
   \vspace{-0.1in}
\end{figure*}

Next, we zoomed into several repositories of each category to examine their activities. 
For each category, we ranked the repositories by the number of stars and selected top (correctly-classified) 100 of them for our study. 
We clone these 600 top repositories and use \emph{git log} command to get the \emph{commit} information about each repository.
Similar to the above analysis on the prevalent creation time of each category ( Figure~\ref{fig:category-create}), we added up the number of commits per day for each category to see their committing trends over time, as shown in Figure~\ref{fig:category-commit}. 
It can be seen that ``Data'' category has far more commits than other categories, which is reasonable because those repositories providing data require constantly updating. 
Besides, the peaks of ``Contact tracing'' and ``Detection and diagnosis'' are somewhat later than others. 
From the above two category-level comparisons we can discover that, \textit{for repositories with diverse aims and focus, their boom period are quite likely to be different}.

\begin{framed}
\noindent \textbf{Answer to RQ2:} 
\textit{
We created a taxonomy and devised an automated approach to classify the repositories into categories.
The \emph{Data} category makes up the largest proportion of them, while the \emph{Contact tracing} category accounts for the least. Subtle differences arise when looking in detail on a per-category basis with regard to the boom time.
}
\end{framed}

\section{Development and maintenance Behaviors}
\label{sec:behavior}

In this section, we direct attention to the
development and maintenance behaviors of these repositories.

\subsection{Contributor Activeness}
Over the past few months, many countries around the world went into lockdown to control the spread of COVID-19. We are interested in investigating whether the quarantine has an impact on the activity of the users on GitHub. 
Our study covers all collected 67,857 contributors.
For each contributor, we collect the number of his/her contributions~(e.g., commits, issues and pull requests) 
to the GitHub repositories in each day of the last year, from August 4, 2019 to August 4, 2020, and use the number of contribution to represent his/her activeness. We analyzed the activity of contributors over the year through two indicators: (a) the total number of active contributors per day (i.e., the number of contributors with contributions greater than 0), and (b) the total number of contributions from all contributors per day.

We can examine the trends of the two indicators over time in three stages, as is shown in Figure~\ref{fig:overall-contributor}. First, before COVID-19 was discovered, they (green and red lines) both exhibit a largely stable trend, with only a slight decline in late-2019 and early-2020, probably due to the Christmas and New Year holidays. Then the latter two stages are prior to and after the WHO's declaration. We can observe a clearly upward trend in both lines when the COVID-19 was beginning to explode globally and the number of confirmed cases rose sharply in March. More specifically, we take two countries, i.e., US and UK, as case studies to analyze whether contributor activity is significantly associated with the corresponding epidemic at the national level. The results are displayed in Figure~\ref{fig:us-contributor} and Figure~\ref{fig:uk-contributor}. We can see in both countries, as the blue line rises sharply, the green line has almost the same upward trend, which is consistent with the overall situation. \textit{These findings suggest that COVID-19 did not significantly disturb contributors' work in GitHub, and may have even given them more time and motivation to contribute.}
Beyond that, it is interesting to see that the trends in both indicators are cyclical and the periodicity is one week upon our inspection.

\begin{figure} [t]
\centering
\subfigure[Commit]{
\label{fig:commit-cluster}
  \includegraphics[width=0.23\textwidth]{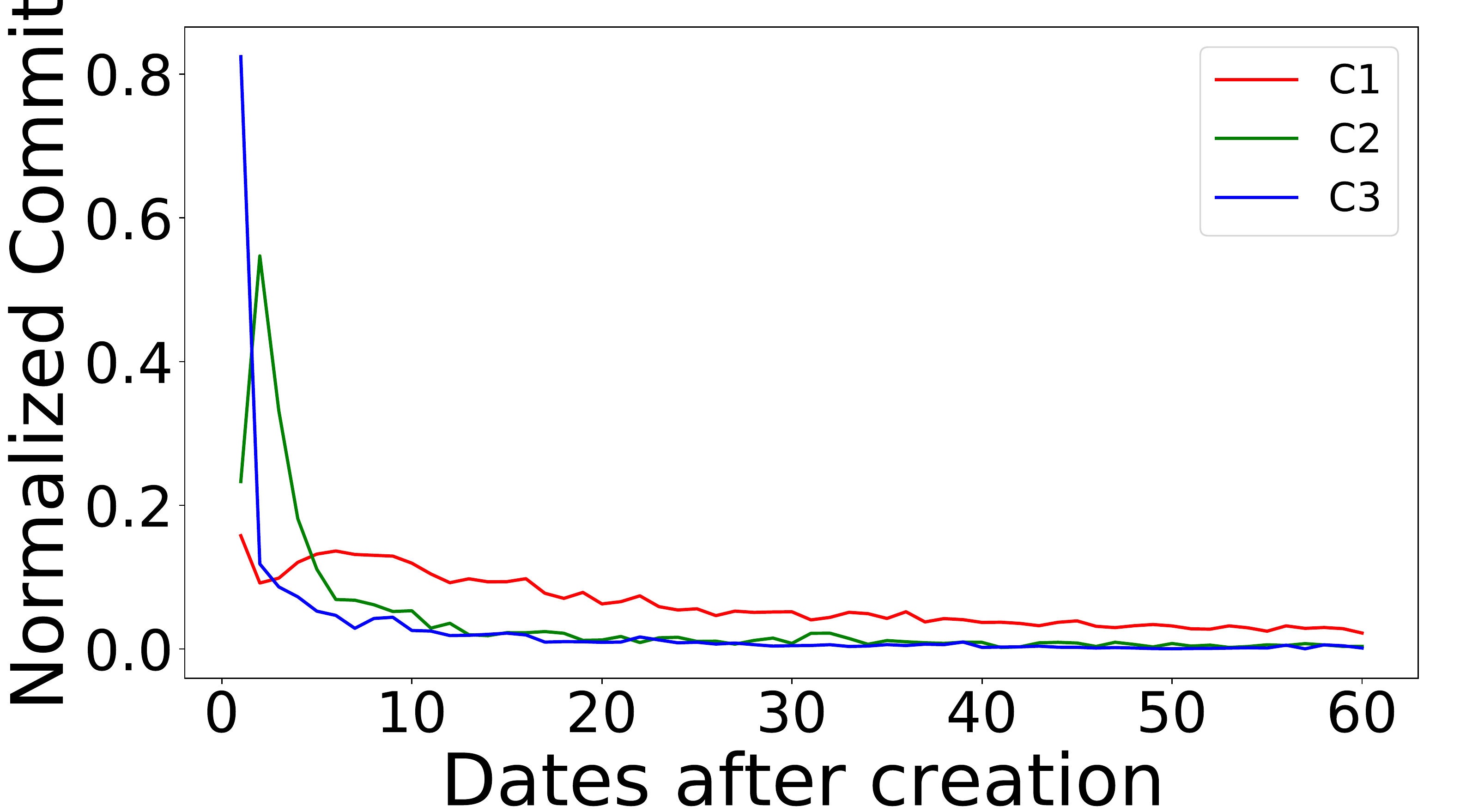}}
\subfigure[Code line]{
\label{fig:codeline-cluster}
  \includegraphics[width=0.23\textwidth]{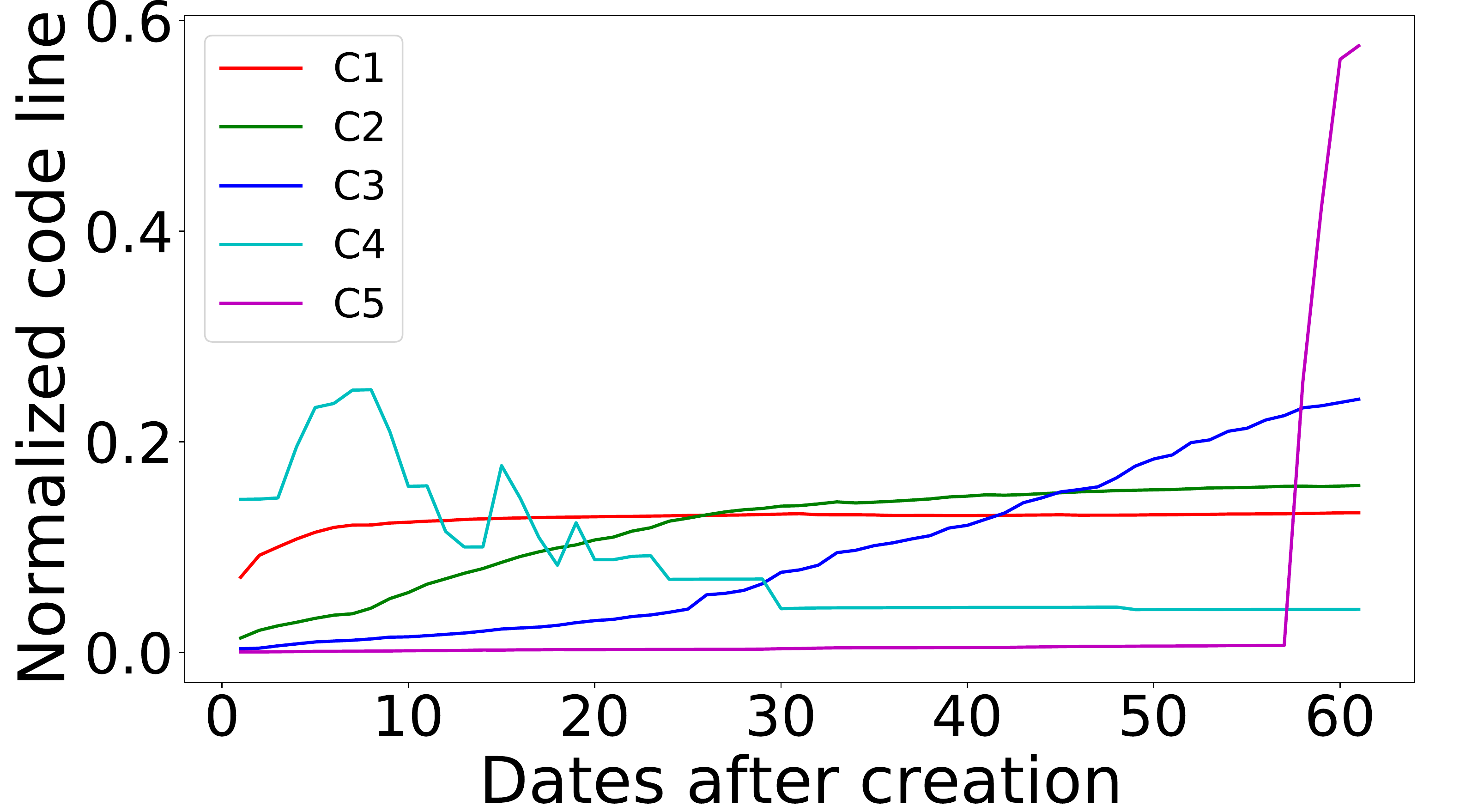}}  
  \vspace{-0.1in}
\caption{The general patterns of \emph{commit} and \emph{line of code}.}
\vspace{-0.2in}
\label{fig:cluster} 
\end{figure}

\subsection{Commit Trends}
We clone the top 100 repositories with the most stars for each category and use \emph{git log} command to get their commit history. 
We are interested in examining how \emph{commit} evolves after a repository is created, so we performed clustering on the postures of \emph{commit} over a period of time of each repository. 

For each repository, we retrieved the number of commits per day over two months (60 days) since it was created, took them as feature vectors and performed L2 regularization. 
Then we applied the K-Means clustering. We used the Elbow Method to find the optimal value of \emph{k}, which is 3.
Thus we grouped the repositories into three clusters. Figure~\ref{fig:commit-cluster} presents the trends of \emph{commit} in each cluster. We can see that \textit{the majority of the commits are done in the first few days after creation}.

1) Cluster 1 (in red) implies a relatively stable trend that with not so many but continuous commits, reflecting the fact that \textit{some repositories have ongoing maintenance after early development}. For example, a repository named ``covid-19-data'' provides data on the COVID-19 confirmed cases around the world. We checked its commit history and found that the \emph{commit} happened almost every day and the number of commits range from 1 to 20 within a day. 

2) Cluster 2 (in green) indicates that the commits are intensive in the first few days after creation, but later the number decreases to nearly 0, which reflects the fact that some repositories are infrequently modified after the initial development is completed. For example, there is a repository called ``covid19'' designed to visualize and track the COVID-19 pandemic by country. It committed frequently in the first five days (average of 27), and then fewer and fewer until 0. 

3) Cluster 3 (in blue) has the similar trend with cluster 2, except that its commits flourish earlier, like the first day when created. For example, there is a repository called ``infectiontracker'', which implements an app for tracing chains of infection and for sharing timely information in the event of COVID-19 infection, with a trend that a lot of commits (93) in the first day, then drops to 0 sharply.

We next investigate the category level distribution (see Table~\ref{tab:cluster-commit}). Most of the repositories in ``Data'' category are clustered into cluster 1, which conforms to Figure~\ref{fig:category-commit}.
Half of the repositories tend to be inactive after a short period of vitality. Given that we choose repositories with a high number of stars, it is reasonable to believe that there exist many more repositories not well maintained for long periods of time.

\begin{table}[t]
\centering
\small
\caption{Distribution of commit clusters in each category}
\vspace{-0.1in}
\resizebox{0.48\textwidth}{!}{

\begin{tabular}{l|c|c|c|c}
\toprule
 Category & \tabincell{c}{ Cluster 1} & \tabincell{c}{Cluster 2} & \tabincell{c}{Cluster 3} & Total \\

\hline
Data & 73 & 16 & 11 & 100 \\
Contact tracing & 34 & 19 & 47 & 100 \\
Toolkit & 44 & 29 & 27 & 100\\
Forecast\&simulation & 56 & 22 & 22 & 100 \\
Detection\&diagnosis & 39 & 29 & 32 &  100 \\
Helpful in some ways & 53 & 25 & 22 & 100 \\
\hline
Total & 299 & 140 & 161 & 600 \\ 
\bottomrule
\end{tabular}
}
\vspace{-0.15in}
\label{tab:cluster-commit} 
\end{table}

\subsection{Lines of Code (LOC)}
We are also interested in exploring the code development process, i.e., the variation of LOC over time. Similar to the analysis of  \emph{commit} evolution, for each repository, we collected the LOC per day over two months (60 days) since it was created, and then applied K-Means clustering likewise.  We also used the Elbow Method to find the optimal value of \emph{k}, which is 5. 
We clustered the data into five clusters and Figure~\ref{fig:codeline-cluster} shows the pattern of each cluster. 

In general, the number of code lines shows an increasing trend, although the period of growth varies. Over 60\% of the repositories (363) are clustered into cluster 1, which shows a tendency of code lines to increase to a certain extent very early on (within five days after creation) and remain stable afterwards, i.e., the code development is concentrated in the very early days and with little or even no change thereafter. For example, there is a repository\footnote{https://github.com/rizmaulana/kotlin-mvvm-covid19} that added 5,398 lines of code on the first day of creation, then gradually increased to 10,182 lines by the 19th day, and has remained stable with handful changes since then. It suggests that most repositories are developed quickly but tend to remain stable after the initial development.
Cluster 2 (164) and cluster 3 (60) manifest a general trend of steady rise, which fits our perception, although the periods of rapid growth are somewhat different and cluster 3 seems to take longer to develop.
Cluster 4 (8) shows that the lines of code fluctuate considerably especially in the first month. To understand this pattern, we manually checked the \emph{commit} messages of several repositories in this cluster. We can observe the following major reasons: \textit{adjust code or data}, \textit{cleanup or refactoring}, \textit{remove data and unused code}, etc. For example, there is a repository\footnote{https://github.com/futurice/corona-simulations} whose lines of code suddenly reduced from 40,343 to 9,926 on the 25th day, and the message of the \emph{commit} is ``\textit{Massive cleanup and refactoring..... Moved all fonts and stylesheets to local. Removed some deprecated code and unneeded data files}''.
Cluster 5 shows that the lines of code remain almost stable over a long period after creation, while increasing sharply at the end time period we focused on, implying that the code development process takes a longer time, like cluster 3. Note that there are only five repositories in cluster 5, indicating that this trend may be very uncommon. We then analyzed the logs of their commits history and found that they added/uploaded files or implemented a new feature on the day the lines of code soared, which are normal occurrences of code development. 
Note that, we further investigate the distribution of different LOC patterns across categories. We observe that they follow the similar distribution with the overall trend, i.e., cluster 1 and 2 are dominant.

\begin{framed}
\noindent \textbf{Answer to RQ3:} 
\textit{
The activity of contributors are not significantly affected by the lockdown of cities. Instead, the quarantine has somewhat boosted contributors' engagement in GitHub. 
Most of the repository development process is prompt and intense in the early stages after creation. Some of the repositories undergo a longer development period (more than two months) and most of them are not well maintained over the long term.   
}
\end{framed}

\section{Discussion}

\subsection{Implications}

Our results show that open source technologies can be rapidly applied to tackle the worldwide public health emergency.
A great many attempts from all over the world have been done on GitHub, covering various aspects from COVID-19 data, computer-aid diagnosis to daily life, and some of them started shortly after the appearance of COVID-19, as well as produced some valuable deliverable with high stars and references.
We notice that contributors' activities in GitHub are not significantly influenced by the lockdown of cities.
Coping with pandemic is an inherently collaborative process. 
The collaboration platform, GitHub, plays an important role to support the software development and information sharing.
It should be noticed that these technologies and resources can be very helpful in other emergencies too.

Our findings also reveal some underlying challenges and propose several potential directions for improvement as follows:
1) A large proportion of repositories are data related, however, GitHub is built on Git and not very suitable to share data.
The platform can be extended or integrate with other data sharing sites for such requirements.
2) Majority of repositories are intra-national and people from different countries may use the native language solely. Internationalization is essential for information exchange in worldwide collaboration. Internationalization techniques can be applied.
3) Most repositories are not well maintained over the long run, and soon become inactive. It is challenging to convene a software project with large number of participants in a short time.
Some mechanisms can be developed for project organization in emergencies.

\subsection{Limitations}
We recognize that our study carries several limitations. First, our investigation is limited by repositories we identified. Although we make efforts to cover all the repositories contains any of the three most representative keywords we summarized, it is quite possible that some COVID-19 related repositories are overlooked by us. Nevertheless, we believe our collection has covered most of the available COVID-19 themed repositories on GitHub. Second, this paper aims to analyze how open source community helps combat COVID-19, however GitHub is not the only platform where people can share their open source projects, which might limit our observations. Third, even though a large number of repositories were collected, only a subset of all COVID-19 relevant repositories were classified and analyzed. The major reason is that most of them are inactive and we cannot understand these repositories based on empty description or description with very few words.

\section{Conclusion}

This paper presents the first large scale empirical study of COVID-19 themed repositories on GitHub. We make efforts to collect over 67K related repositories and characterize them from trend, popularity, contributors, etc. To further understand the development and maintenance behaviors of COVID-19 themed repositories, we propose a NLP-based method to classify them and then perform a per category analysis. Our observations show the promising direction of applying open source technologies to tackle public health emergency, and reveal some underlying challenges for improvement.

\balance

\bibliographystyle{IEEEtran}
\bibliography{cite}

\end{document}